\documentclass[sigconf]{acmart}
\AtBeginDocument{%
  \providecommand\BibTeX{{%
    \normalfont B\kern-0.5em{\scshape i\kern-0.25em b}\kern-0.8em\TeX}}}

\setcopyright{acmcopyright}
\copyrightyear{2022}
\acmYear{2022}
\acmDOI{}

\acmConference[CIKM '22]{Make sure to enter the correct
  conference title from your rights confirmation emai}{October 17--22,
  2022}{Atlanta, GA}
\acmPrice{}
\acmISBN{}

\usepackage{graphicx}
\usepackage{caption}
\usepackage{subcaption}
\newcommand{\med}{\mbox{RetroDCVAE}}
\usepackage{tabularx}
\usepackage{booktabs}
\usepackage{multirow}
\usepackage{multicol}
\usepackage{adjustbox}
\usepackage{bm}
\DeclareMathOperator*{\argmax}{arg\,max}

\usepackage{pifont}
\newcommand{\cmark}{\ding{51}}
\newcommand{\xmark}{\ding{55}}
\usepackage{enumitem}
\usepackage{xcolor}

\begin{document}

% \title{RetroDCVAE: Generating Diverse Reactants for  Retrosynthesis via Discrete Conditional Variational Autoencoders}
\title{Modeling Diverse Chemical Reactions for Single-step Retrosynthesis via Discrete Latent Variables}
\author{Huarui He}
\affiliation{
    \institution{CAS Key Laboratory of Technology in GIPAS,}
    \institution{University of Science and Technology of China,}
    \city{Hefei}\country{China}}
\email{huaruihe@mail.ustc.edu.cn}
\author{Jie Wang}
\affiliation{
    \institution{CAS Key Laboratory of Technology in GIPAS,}
    \institution{University of Science and Technology of China,}
    \institution{Institute of Artificial Intelligence,
Hefei Comprehensive National Science Center,}
    \city{Hefei}\country{China}}
\email{jiewangx@ustc.edu.cn}
\authornote{Corresponding Author}
\author{Yunfei Liu}
\affiliation{
    \institution{CAS Key Laboratory of Technology in GIPAS,}
    \institution{University of Science and Technology of China,}
    \city{Hefei}\country{China}}
\email{yf1274540173@mail.ustc.edu.cn}
\author{Feng Wu}
\affiliation{
    \institution{CAS Key Laboratory of Technology in GIPAS,}
    \institution{University of Science and Technology of China,}
    \city{Hefei}\country{China}}
\email{fengwu@ustc.edu.cn}

\renewcommand{\shortauthors}{Anon.}

\begin{abstract}
Single-step retrosynthesis is the cornerstone of retrosynthesis planning, which is a crucial task for computer-aided drug discovery.
The goal of single-step retrosynthesis is to identify the possible reactants that lead to the synthesis of the target product in one reaction.
By representing organic molecules as canonical strings, existing sequence-based retrosynthetic methods treat the product-to-reactant retrosynthesis as a sequence-to-sequence translation problem.
However, most of them struggle to identify diverse chemical reactions for a desired product due to the deterministic inference, which contradicts the fact that many compounds can be synthesized through various reaction types with different sets of reactants.

In this work, we aim to increase reaction diversity and generate various reactants using discrete latent variables.
We propose a novel sequence-based approach, namely \med, which incorporates conditional variational autoencoders into single-step retrosynthesis and associates discrete latent variables with the generation process.
Specifically, \med~uses the Gumbel-Softmax distribution to approximate the categorical distribution over potential reactions and generates multiple sets of reactants with the variational decoder.
Experiments demonstrate that \med~outperforms state-of-the-art baselines on both benchmark dataset and homemade dataset.
Both quantitative and qualitative results show that \med~can model the multi-modal distribution over reaction types and produce diverse reactant candidates.
% \footnote{Our code and homemade dataset are publicly available at \url{https://github.com/anonymous2262/RetroDCVAE}.}

% As we can represent organic molecules as canonical strings, the product-to-reactant retrosynthesis naturally turns into a sequence-to-sequence translation problem.
% %
% However, due to the deterministic inference, many existing sequence-based retrosynthetic methods struggle to model the \textit{one-to-many} retrosynthesis scenario, where we can synthesize a desired product by different sets of reactants.
\end{abstract}






\maketitle

\section{Introduction}

As a fundamental problem in organic chemistry, retrosynthesis planning refers to the technique to iteratively deconstruct a compound into intermediates or simpler precursors, until a set of commercially available reactants is reached.
Since we can solve retrosynthesis planning by searching backwards and recursively applying single-step retrosynthesis to unavailable molecules, this work focuses on single-step retrosynthesis whose goal is to generate a set of reactants that leads to the one-step synthesis of the desired product.
% We focus on single-step retrosynthesis as retrosynthesis planning can be solved by searching backwards and recursively applying single-step retrosynthesis to unavailable molecules.
In recent years, we have witnessed the great achievement of computer-aided synthesis planning (CASP) \cite{corey1969computer} in many real-world applications, such as drug design \cite{retrognn}, environmental protection \cite{trost1991atom}, as well as materials science \cite{yuan2018retrosynthesis}. 

\begin{figure}
  \includegraphics[width=\linewidth]{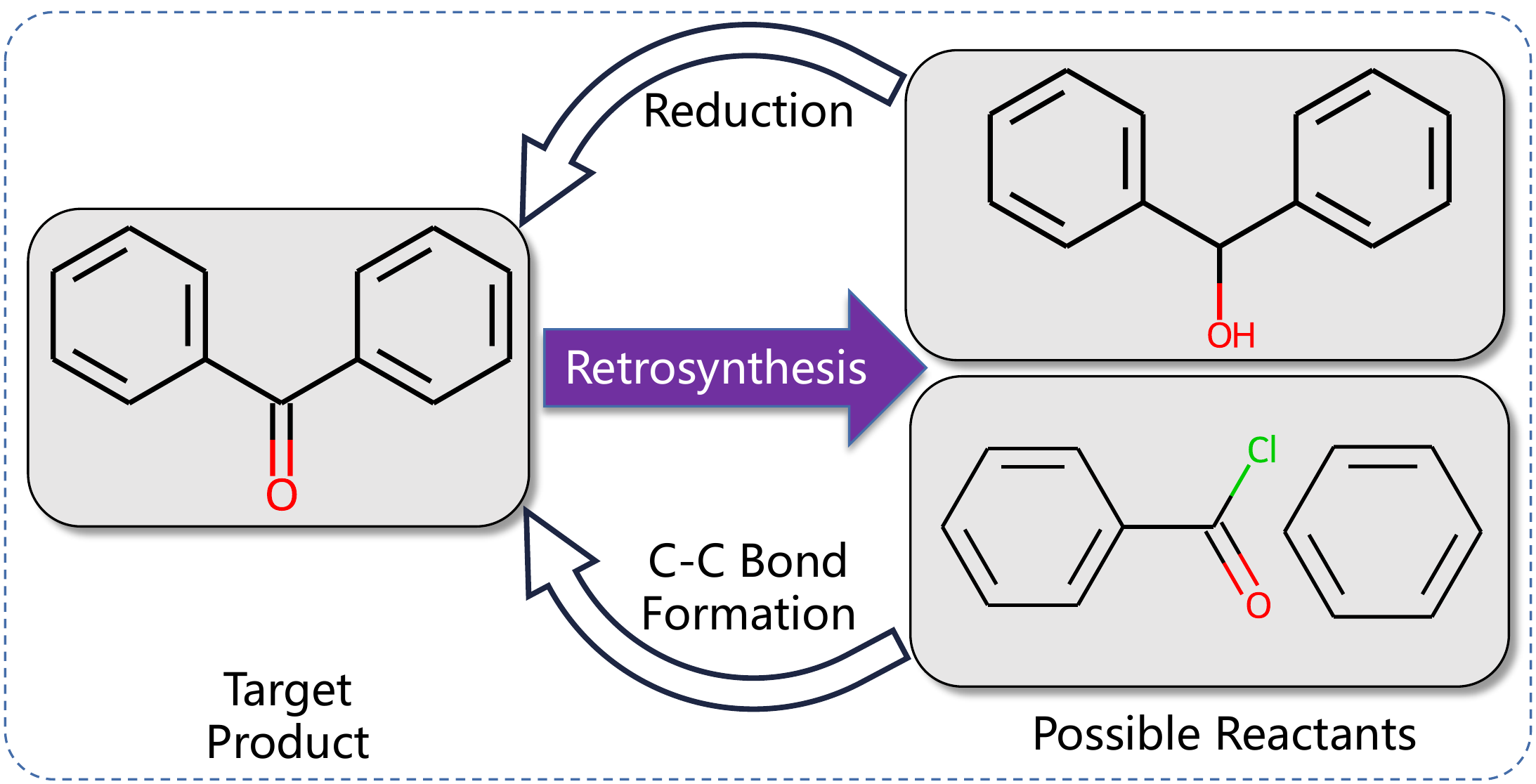}
\caption{Illustration of single-step retrosynthesis with two possible reactions. We can synthesize the target product Benzophenone using different sets of reactants through different reactions. Either of the synthesis schemes is feasible. Specifically, Benzophenone (Left) can be synthesized by Benzhydrol through Reduction (Top) or by Benzoyl chloride and Benzene through C-C bond formation (Bottom).}
% product 二苯甲酮，第二个是二苯甲醇，第三个是苯甲酰氯
\label{fig:motivation}
\vspace{-3.5mm}
\end{figure}

Given that the simplified molecular-input line-entry system (SMILES) \cite{weininger1988smiles} can represent arbitrary molecules as strings, researchers formulate the single-step retrosynthesis task as a machine translation problem \cite{Lin_2020_AutoSynRoute, Zheng_2020_SCROP, Tetko_2020_AT}. 
Therefore, sequence-based approaches in natural language processing \cite{seq_to_seq, vaswani2017attention} lead to a simple end-to-end formulation that exempts retrosynthesis from reaction templates and domain knowledge.
As the extract of high-quality reaction templates requires considerable experience and expertise, the sequence-based retrosynthesis has attracted increasing attention and shown promise as a prevailing template-free approach.

However, existing work \cite{Lin_2020_AutoSynRoute, Zheng_2020_SCROP, Tetko_2020_AT, g2s} that adapts sequence-to-sequence models to retrosynthesis struggles to model the multi-modal distribution over reaction types.
For example, Figure \ref{fig:motivation} illustrates that Benzophenone can be synthesized in one step by the Reduction reaction or by the C-C bond formation reaction.
As existing template-free retrosynthetic models mainly perform deterministic inference, once we finish training and fix the parameters, the retrosynthetic models tend to generate reactants in one direction.
That is, existing template-free models may always output only one possible synthesis scheme due to the deterministic inference.

In this paper, we propose a novel template-free model, namely \med, which incorporates stochastic inference into single-step retrosynthesis.
Motivated by the fact that a large number of compounds have multiple synthetic schemes, we implicitly model reaction types with categorical latent variables.
However, neural networks with discrete latent variables are challenging to train due to the inability to backpropagate through samples.
Built upon the Transformer architecture \cite{vaswani2017attention}, \med~incorporates conditional variational autoencoders (CVAE) \cite{cvae} into the Transformer decoder part and uses the Gumbel-Softmax estimator \cite{gumbel} to support backpropagation of categorical latent variables.
To gain insights into the working behaviors of \med, we conduct analysis on a homemade dataset, namely USPTO-DIVERSE, where each target compound owns at least two sets of reactants.
Both quantitative and qualitative results show that \med~is able to generate diverse reactants for a target product.
Furthermore, experiments on the public dataset demonstrate that \med~achieves competitive results against state-of-the-art baselines.

The contributions of this work are threefold:

\noindent (1) To the best of our knowledge, we are the first to model the multi-modal distribution in single-step retrosynthesis using discrete CVAE. Moreover, we leverage Gumbel-Softmax \cite{gumbel} to optimize the discrete CVAE for the first time.

\noindent (2) To assess the capacity of modeling the multi-modal distribution over reaction types, we extract all reactions with 1-to-N $(N\ge 2)$ products in USPTO-MIT \cite{jin2017predicting} to build a new dataset named USPTO-DIVERSE. That is, each product in USPTO-DIVERSE corresponds to two or more reactions. We anticipate that USPTO-DIVERSE would exert potential impacts on community.

% select all 1-to-N $(N\ge 2)$ reactions in USPTO-MIT \cite{jin2017predicting} to build a new dataset named USPTO-DIVERSE.
% homemade dataset
% To gain insights into the working behaviors of \med, we conduct analysis on USPTO-DIVERSE, which consists of all 1-to-N $(N\ge 2)$ reactions in USPTO-MIT \cite{jin2017predicting}.

\noindent (3) We conduct both quantitative and qualitative analysis to show the effectiveness of the proposed \med. Extensive experiments demonstrate that \med~outperforms state-of-the-art template-free baselines on the benchmark dataset USPTO-50k and the homemade dataset USPTO-DIVERSE.

\section{Preliminaries}
% \section{Background}
% \hspace{1.5mm}
In this section, we first formulate the single-step retrosynthetic problem and then introduce basic techniques for multi-modal distribution modeling.
\subsection{Problem Statement}
In this part, we define the single-step retrosynthetic task and present the notations throughout this paper.

Let $\mathcal{M}$ denote the space of all molecules and $\mathcal{R}$ denote the space of all chemical reaction types.
The single-step retrosynthetic model takes a target molecule $t\in \mathcal{M}$ as input, and predicts a set of precursor source reactants $\mathcal{S}\subset \mathcal{M}$ leading to synthesizing $t$. 
Therefore, single-step retrosynthesis is the reverse problem of reaction outcome prediction, whose goal is to predict the major product given the set of reactants. 
Admittedly, single-step retrosynthesis is more challenging than its reverse problem due to the multi-modal distribution over reaction types for a given product.
Formally, a single-step retrosynthetic model aims to find a function $F$ as
\[
F(\cdot): \quad t\rightarrow \{\mathcal{S}_i\}_{i=1}^b,
\]
which outputs at most $b$ sets of reactants.
We expect that $F$ ranks ground truth or plausible reactants as high as possible.
The single-step retrosynthetic model can be learned from the collection of chemical reactions $\mathcal{D}_{\text{train}}=\{\mathcal{S}_i, t_i \}$ in patents granted by United States Patent Office (USPTO).

From a sequence-based perspective, we use SMILES to represent the target product $t$ as a string. For example, we represent Benzophenone as O=C(C1C=CC=CC=1)C1C=CC=CC=1.
Existing sequence-based retrosynthetic models employ the encoder-decoder architecture.
The encoder embeds each token of $t$ into the continuous vector space and drives a sequence of embeddings as $\mathbf{x}=\{\mathbf{x}_1,\cdots, \mathbf{x}_{T_i}\}$ with ${T_i}$ representing the length of the input SMILES string. 
After that, instead of directly performing deterministic inference and decoding $\mathbf{x}$, we aim to implicitly model possible reaction types from $\mathbf{x}$ and then perform stochastic inference. Given $\mathbf{x}$, the multi-modal distribution over reaction types is $\textup{P}(\mathbf{z}|\mathbf{x})$. During the inference phase, we sample a possible reaction type $\mathbf{z}\in \mathcal{R}$ following the distribution $\mathbf{z}\sim \textup{P}(\mathbf{z}|\mathbf{x})$.
According to $\{\mathbf{x},\mathbf{z}\}$, the auto-regressive decoder generates a sequence of chemical symbol predictions $\mathbf{y}=\{\mathbf{y}_1, \cdots, \mathbf{y}_{T_o}\}$ with ${T_o}$ representing the length of output tokens. 
Finally, \textit{beam search} is applied to derive the string representation of reactants.
For example, if the sampled $\mathbf{z}\in \mathcal{R}$ means C-C bond formation reaction, then the retrosynthetic model aims to synthesize Benzophenone through C-C bond formation reaction and would output C1=CC=CC=C1.C1=CC=CC(C(Cl)=O)=C1 as the final prediction, where ``.'' is the delimiter of two molecules.

\subsection{Technical Background}
\textbf{CVAE} allows us to tackle problems where the input-to-output mapping is one-to-many, while reducing the need to explicitly specify the structure of the output distribution.
CVAE \cite{cvae} aims to model the underlying conditional distribution $p(\mathbf{y}|\mathbf{x})$, i.e., the distribution over output variable $\mathbf{y}$, conditioned on the observed variable $\mathbf{x}$.
Given observation $\mathbf{x}$, the latent variable $\mathbf{z}$ is drawn from the prior distribution $p(\mathbf{z}|\mathbf{x})$, and the output $\mathbf{y}$ is generated from the distribution $p(\mathbf{y}|\mathbf{x},\mathbf{z})$.
By assuming $\mathbf{z}$ follows multivariate Gaussian distribution with a diagonal co-variance matrix, we have
\begin{equation}\label{eqn:integ}
p(\mathbf{y}|\mathbf{x})
=\int_\mathbf{z} p(\mathbf{y}|\mathbf{x},\mathbf{z})p(\mathbf{z}|\mathbf{x})d\mathbf{z}.
\end{equation}
The typical CVAE consists of a prior network $p_\theta (\textbf{z}|\textbf{x})$ parameterized by $\theta$ and a recognition network $q_\phi (\mathbf{z}|\mathbf{x},\mathbf{y})$ parameterized by $\phi$, which are used to approximate the prior distribution $p(\textbf{z}|\textbf{x})$ and the posterior distribution $p(\mathbf{z}|\mathbf{x},\mathbf{y})$, respectively.
The evidence lower bound (ELBO) is
\begin{equation}\label{eqn:elbo}
% \begin{align*}
\begin{split}
    \mathcal{L}_{\text{ELBO}} 
    &=\mathcal{L}_{\text{REC}}-\mathcal{L}_{\text{KL}} \\
    &=\mathbb{E}_{q_\phi(\mathbf{z}|\mathbf{x},\mathbf{y})}[\log p_\theta(\mathbf{y}|\mathbf{x},\mathbf{z})]-D_{\text{KL}}(q_\phi(\mathbf{z}|\mathbf{x},\mathbf{y})\|p_\theta (\textbf{z}|\textbf{x})) \\
    &\leq \log p_\theta(\mathbf{y}|\mathbf{x}),
% \end{align*}
\end{split}
\end{equation}
where $\mathcal{L}_{\text{REC}}$ is the negative reconstruction error and $\mathcal{L}_{\text{KL}}$ denotes the Kullback-Leibler (KL) divergence between the posterior and prior.
Therefore, to maximize the ELBO w.r.t. parameters $\theta$ and $\phi$ will concurrently maximize the likelihood $p(\mathbf{y}|\mathbf{x})$ and minimize the KL divergence.

\noindent\textbf{Gumbel-Softmax} is a continuous distribution that can approximate categorical samples. We can easily compute parameter gradients of Gumbel-Softmax via the reparameterization trick \cite{vae,DBLP:conf/icml/RezendeMW14}. 
Let $\mathbf{z}$ be a categorical variable with class probabilities $\pi_1, \pi_2,\cdots,\pi_k$.
To draw samples $\mathbf{z}$ from a categorical distribution with class probabilities $\mathbf{\pi}$, \citet{gumbel1954statistical} and \citet{maddison2014sampling} propose the Gumbel-Max trick, i.e.,
\[
\mathbf{z}=\text{one\_hot}\big(\argmax_{i\in [k]} (g_i+\log \pi_i) \big),
\]
where $[k]=\{1,2,\cdots,k\}$ and samples $g_1,\cdots, g_k$ are drawn i.i.d. according to $\text{Gumbel}(0,1)$. In practice, the Gumbel$(0,1)$ distribution can be derived using inverse transform sampling by drawing $u\sim \text{Uniform}(0,1)$ and computing $g=-\log (-\log u)$.

However, the $\argmax$ formulation makes it hard to backpropagate through samples. \citet{gumbel} use the softmax function as a continuous and differentiable approximation to $\argmax$, i.e.,
\[
\mathbf{z}_i = \frac{\exp((\log\pi_i + g_i)/\tau)}{\sum_{j=1}^k \exp((\log\pi_j+g_j)/\tau)}, \quad\text{for }i\in [k].
\]
As the softmax temperature $\tau$ approaches 0, \citet{gumbel} point out that samples from the Gumbel-Softmax distribution become one-hot and the Gumbel-Softmax distribution becomes identical to the categorical distribution $p(\mathbf{z})$.

\begin{figure}[t]
\centering % <-- added
\includegraphics[width=\linewidth]{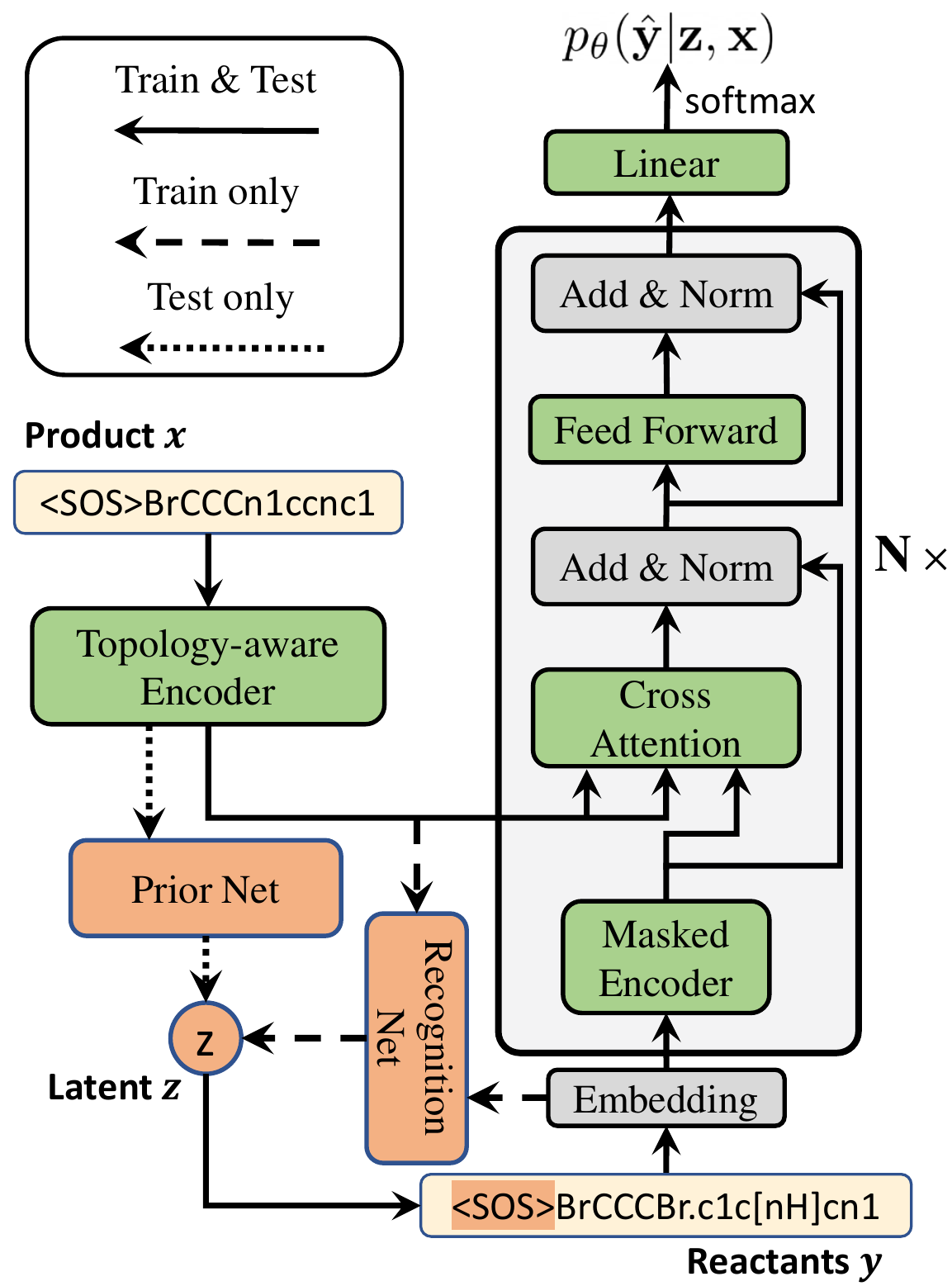}
\caption{Illustration of proposed \med.}
\label{fig:pipeline}
\end{figure}

\section{Methodology} \label{sec:med}
The proposed discrete conditional stochastic inference mechanism is widely applicable to existing sequence-based models.
As shown in Figure \ref{fig:pipeline}, we follow the encoder-decoder architecture and incorporate the stochastic inference mechanism into the decoder part.
As for the encoder part, we perform directed message passing and follow a state-of-the-art template-free baseline \cite{g2s} to convert the input product SMILES into a sequence of topology-aware embeddings $\mathbf{x}$. 
Given $\mathbf{x}$, we develop a variational auto-regressive decoder to generate the output sequence of chemical symbol predictions $\mathbf{y}$. 
In this section, we first briefly introduce the topology-aware encoder in Section \ref{sub:enc}. 
After that, we elaborate the proposed variational auto-regressive decoder in Section \ref{sub:dec}.

\subsection{Topology-aware Encoder}\label{sub:enc}
We can represent molecules in multiple ways, such as molecular fingerprints \cite{DBLP:journals/jcisd/RogersH10}, SMILES strings \cite{weininger1988smiles}, and molecular graphs with atoms as nodes and bonds as edges.
As \citet{jin2018junction} criticize that the linear SMILES representation can not well represent the topological context of atoms in a molecular graph, we embed molecules into the continuous vector space as graph embeddings via a topology-aware molecular graph encoder. 
In the following, we introduce the two critical components of the topology-aware encoder, namely directed message passing neural network (D-MPNN) \cite{dmpnn} and topology-aware positional embedding \cite{g2s}.

% \noindent\textbf{D-MPNN.} 
\subsubsection{D-MPNN} Unlike atom-oriented message passing in edge-aware MPNNs \cite{gnn-ptm,Mao_2021_GET}, we follow D-MPNN \cite{dmpnn} and use a Directed Graph Convolutional Network (D-GCN) to derive topology-aware atom representations, where message updates are oriented towards directed bonds to prevent totters, or messages being passed back-and-forth between neighbors \cite{dmpnn}.
Specifically, suppose the message passing performed on an undirected molecular graph $\mathcal{G}$ with atom features $x_v$ and bond features $e_{vw}$ consists of $T$ steps. On each step $t$, hidden states $h_{vw}^{(t)}$ and messages $m_{vw}^{(t)}$ associated with each vertex $v$ are updated through aggregate function $M^{(t)}$ and update function $U^{(t)}$ as
% The corresponding message passing process can be written as 
\begin{align*}
    m_{vw}^{(t+1)} &=\sum_{k\in \{\mathcal{N}(v)\setminus w\}}M^{(t)}(x_v, x_k, h_{kv}^{(t)}),\\
    h_{vw}^{(t+1)} &=U^{(t)}(h_{vw}^{t}, m_{vw}^{(t+1)}),
\end{align*}
where $\mathcal{N}(v)$ is the set of neighboring atoms of $v$ in the molecular graph $\mathcal{G}$. Note that the direction of messages matters since we initialize edge hidden states as $h_{vw}^{(0)}=\textup{ReLU}(W [x_v \| e_{vw}])$ with the trainable matrix $W$. After $T$ iterations, we obtain final atom representations by
\begin{align*}
    m_v &= \sum_{w\in \mathcal{N}(v)}h_{wv}^{(T)},\\
    h_v &= \textup{GELU}(W_o[x_v \| m_v]),
\end{align*}
where $W_o$ is a trainable matrix, $\textup{GELU}$ denotes the GELU activation \cite{gelu}, and $\|$ denotes the concatenation operation.

\subsubsection{Topology-aware token embedding.} To capture atom interactions, the atom representations coming out of D-MPNN are fed into Transformer encoders.
Inspired by the relative positional embedding used in Transformer-XL \cite{transformer-xl}, we follow Graph2SMILES \cite{g2s} and use the topology-aware positional embedding, which is dependent on the shortest path length between atoms in a molecular.
Specifically, the attention score between atoms $v$ and $w$ in the standard Transformer \cite{vaswani2017attention} can be decomposed as \begin{align*}
    \mathbf{A}_{v,w}^{abs} &=
\mathbf{h}_v^\top \mathbf{W}_q^\top \mathbf{W}_k \mathbf{h}_w
+\mathbf{h}_v^\top\mathbf{W}_q^\top\mathbf{W}_k \mathbf{p}_w \\
&+\mathbf{p}_v^\top \mathbf{W}_q^\top \mathbf{W}_k \mathbf{h}_w
+\mathbf{p}_v^\top\mathbf{W}_q^\top\mathbf{W}_k \mathbf{p}_w,
\end{align*}
where $\mathbf{W}_q$, $\mathbf{W}_k$ are weights for the keys and queries, and $\mathbf{p}_v$, $\mathbf{p}_w$ are the absolute positional embeddings corresponding to atoms $v$, $w$.
Similar to Transformer-XL \cite{transformer-xl}, we follow Graph2SMILES \cite{g2s} and reparameterize the four terms as follows,
\begin{align*}
\mathbf{A}_{v,w}^{rel} &=
(\mathbf{h}_v^\top \mathbf{W}_q^\top +\mathbf{c}^\top ) \mathbf{W}_k \mathbf{h}_w
+(\mathbf{h}_v^\top \mathbf{W}_q^\top +\mathbf{d}^\top ) \mathbf{W}_{k,R} \mathbf{r}_{v,w} \\
&=(\mathbf{h}_v^\top \mathbf{W}_q^\top +\mathbf{c}^\top ) \mathbf{W}_k \mathbf{h}_w
+(\mathbf{h}_v^\top \mathbf{W}_q^\top +\mathbf{d}^\top ) \tilde{\mathbf{r}}_{v,w},
\end{align*}
where the trainable biases are renamed as $\mathbf{c}$ and $\mathbf{d}$ to avoid confusion and $\mathbf{r}_{v,w}$ is the learnable embedding dependent on the shortest path length between atoms $v$ and $w$.
For convenience, we merge $\mathbf{W}_{k,R} \mathbf{r}_{v,w}$ into $\tilde{\mathbf{r}}_{v,w}$ and use the shortest path length between atoms $v$ and $w$ to look up $\tilde{\mathbf{r}}_{v,w}$ in the trainable positional embedding matrix.
Therefore, our encoder integrates topological context into the final token embeddings of the input molecule.

\subsection{Variational Auto-regressive Decoder}\label{sub:dec}
\begin{figure}[!t]
\centering % <-- added
\includegraphics[width=0.85\linewidth]{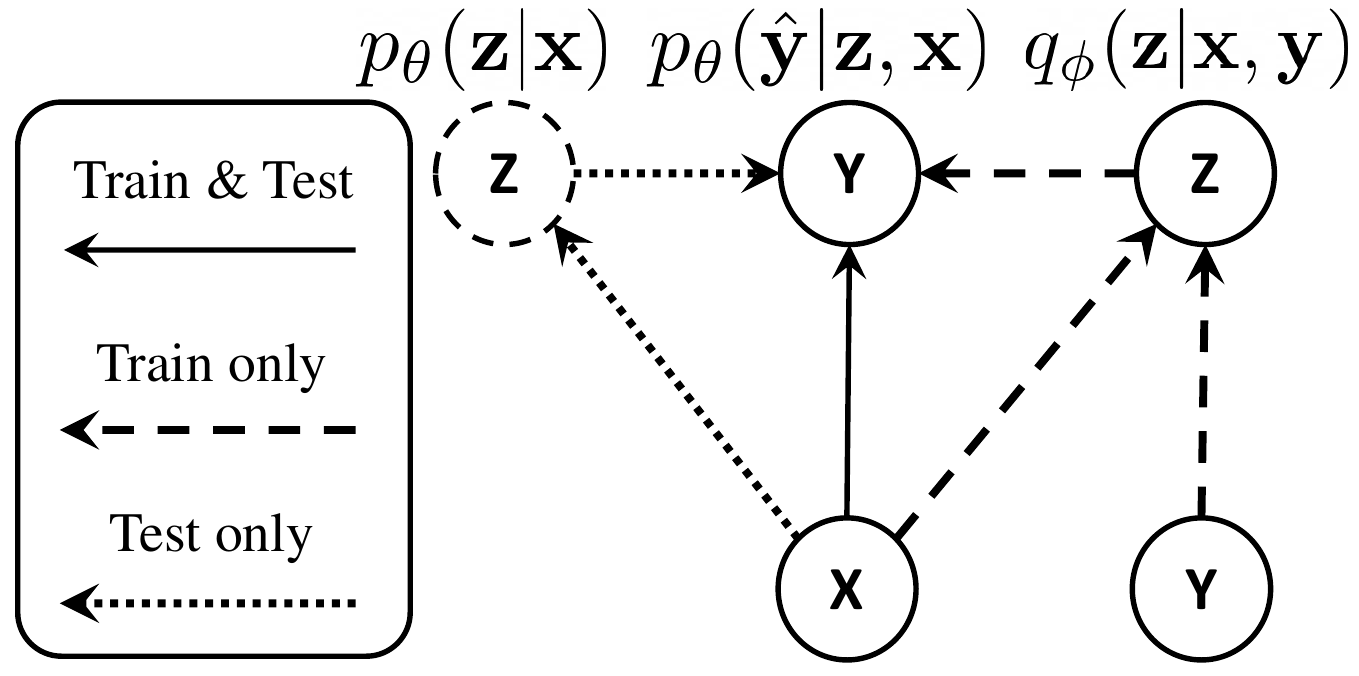}
\caption{Conditional directed graphical model.}
\label{fig:cgm}
\vspace{-2mm}
\end{figure}

We consider the decoding process as a conditional graphical model concerning three random variables: the encoder output $\mathbf{x}$, the discrete latent variable $\mathbf{z}$, and the target reactants $\mathbf{y}$.
Figure \ref{fig:cgm} illustrates the conditional directed graphical model, where the encoder output is viewed as the observed context.
For a given observation $\mathbf{x}$, the latent variable $\mathbf{z}$ is drawn from the prior distribution $p_\theta (\mathbf{z}|\mathbf{x})$, 
and the decoder output $\hat{\mathbf{y}}$ is generated from the conditional distribution 
$p_\theta (\hat{\mathbf{y}}|\mathbf{x}, \mathbf{z})$.
We interpret the one-hot latent variable $\mathbf{z}$ as the potential reaction type which directs the reactant generating process.

Our variational auto-regressive decoder (VAD) is trained to maximize the conditional log-likelihood $\log p_\theta (\mathbf{y}|\mathbf{x})$. As the objective function in Equation \ref{eqn:integ} is intractable, we use the variational lower bound of the log-likelihood in Equation \ref{eqn:elbo} as a surrogate objective function.
Under the reparameterization \cite{vae}, we can replace an expectation w.r.t. $q_\phi(\mathbf{z}|\mathbf{x},\mathbf{y})$ with one w.r.t. $p(\mathbf{\epsilon})$. Therefore, we rewrite the ELBO in Equation \ref{eqn:elbo} as
\begin{align*}
\mathcal{L}_{\theta,\phi} (\mathbf{x},\mathbf{y})=
\mathbb{E}_{p(\mathbf{\epsilon})}\left[\log p_\theta (\mathbf{y}|\mathbf{x}, \mathbf{z}) \right] -D_{\text{KL}}\left(q_\phi(\mathbf{z}|\mathbf{x},\mathbf{y})\|p_\theta (\mathbf{z}|\mathbf{x}) \right),
\end{align*}
where $\mathbf{z}=g_\phi (\mathbf{x},\mathbf{y},\mathbf{\epsilon})$, with $g_\phi(\cdot,\cdot, 
\cdot)$ indicating a deterministic, differentiable function and the distribution of the random variable $\mathbf{\epsilon}$ is independent of $\mathbf{x}$ or $\mathbf{y}$. As a result, we can form a simple Monte Carlo estimator $\Tilde{\mathcal{L}}_{\theta,\phi} (\mathbf{x},\mathbf{y})$ of the ELBO as an empirical lower bound where we use a single noise sample $\mathbf{\epsilon}$ from the Gumbel distribution, i.e.,
\begin{align*}
    \mathbf{\epsilon}&\sim \text{Gumbel}(0,1),\\
    \mathbf{z}&=g_\phi (\mathbf{x},\mathbf{y},\mathbf{\epsilon}),\\
    \Tilde{\mathcal{L}}_{\theta,\phi} (\mathbf{x},\mathbf{y})&=\log p_\theta (\mathbf{y}|\mathbf{x}, \mathbf{z})-D_{\text{KL}}\left(q_\phi(\mathbf{z}|\mathbf{x},\mathbf{y})\|p_\theta (\mathbf{z}|\mathbf{x}) \right).
\end{align*}

\noindent\textbf{Inference process.}
Variational inference is a technique for approximating an intractable posterior distribution  with a tractable surrogate. Figure \ref{fig:cgm} illustrates that $p(\mathbf{z}|\mathbf{x}, \mathbf{y})$ and $p (\mathbf{z}|\mathbf{x})$ are approximated with a recognition model and a prior model, respectively.
Instead of separately and iteratively optimizing the variational parameters per datapoint, we use the same recognition network (with parameters $\phi$) and prior network (with parameters $\theta$) to perform posterior inference over all of the samples $(\mathbf{x}, \mathbf{y})$ in our dataset.
The strategy of sharing variational parameters across datapoints is also known as amortized variational inference \cite{avi}.
With amortized inference, \med~can avoid a per-datapoint optimization loop and leverage the efficiency of stochastic gradient descent.

\noindent\textbf{Parameter learning.}
We summarize the pipeline of variational decoding in Figure \ref{fig:pipeline}. 
Both recognition network $q_\phi(\mathbf{z}|\mathbf{x},\mathbf{y})$ and (conditional) prior network $p_\theta (\mathbf{z}|\mathbf{x})$ are implemented with GRUs \cite{gru} as both $\mathbf{x}$ and $\mathbf{y}$ are time series data.
Taking the prior network $p_\theta (\mathbf{z}|\mathbf{x})$ for an example, we can formulate the workflow at step $t$ as 
\begin{align*}
    \mathbf{s}^{(t)}&=\sigma(\mathbf{W}_s\cdot\mathbf{x}^{(t)}+\mathbf{U}_s\cdot \mathbf{h}^{(t-1)}+\mathbf{b}_s),\\
    \mathbf{r}^{(t)}&=\sigma(\mathbf{W}_r\cdot \mathbf{x}^{(t)}+\mathbf{U}_r\cdot \mathbf{h}^{(t-1)}+\mathbf{b}_r),\\
    \Tilde{\mathbf{h}}^{(t)}&=\text{tanh}(\mathbf{W}\cdot\mathbf{x}^{(t)}+\mathbf{U}\cdot(\mathbf{r}^{(t)}\odot\mathbf{h}^{(t-1)})+\mathbf{b}),\\
    \mathbf{h}^{(t)}&=(1-\mathbf{s}^{(t)})\odot\mathbf{h}^{(t-1)}+\mathbf{s}^{(t)}\odot\Tilde{\mathbf{h}}^{(t)},
\end{align*}
where $\odot$ means the element-wise multiplication.
For convenience, we summarize the process as
\[
\mathbf{h}^{(t)}=\text{GRU}_\theta(\mathbf{x}^{(t)}, \mathbf{h}^{(t-1)}), \quad \text{for }t\in [T_i],
\]
where $\mathbf{h}^{(0)}$ is initialized to zero vector.

As reaction types are discrete, we employ the Gumbel-Softmax distribution to approximate the latent categorical variable, i.e.,
\begin{align*}
    \mathbf{\pi}_j&=\frac{\exp(\mathbf{h}^{(T_i)}_j)}{\sum_{k=1}^K \exp(\mathbf{h}^{(T_i)}_k)},\\
    \mathbf{z}_j&=\frac{\exp((\log\pi_j + g_j)/\tau)}{\sum_{k=1}^K \exp((\log\pi_k+g_k)/\tau)}, \quad\text{for }j\in [K],
\end{align*}
where $g_1,\cdots, g_K$ are independently drawn from $\text{Gumbel}(0,1)$ and $K$ is the predefined latent size.
Throughout this paper, we further discretize $\mathbf{z}$ with $\argmax$ but use the continuous approximation in the backward pass, which is known as Straight-Through Gumbel Estimator \cite{gumbel}.
As depicted in Figure \ref{fig:pipeline}, our VAD initializes the \textit{start of the sequence} (namely the ``\textit{<SOS>}'' token) based on $\mathbf{z}$ and then decodes subsequent tokens autoregressively.

\section{Experimental Setup} \label{sub:setup}
In this section, we detail the experimental setup including datasets, metrics, baselines, and empirical tricks.
\subsection{Datasets}
We empirically evaluate our model on both public and homemade datasets derived from granted patents.
We reserve all reactions containing 1-to-N products in USPTO-MIT \cite{jin2017predicting} to create USPTO-DIVERSE, where the maximum $N$ is 39 in the training set, i.e., there are 39 reactions in the training set that share the same product.
% 39 reactions in the training set share the same product.
In Section \ref{sub:uspto-diverse}, we provide both quantitative and qualitative results on homemade USPTO-DIVERSE to prove the claim that \med~is able to improve the performance of 1-to-N retrosynthesis and generate diverse reactants.
In Section \ref{sub:uspto-50k}, we compare the proposed approach with existing single-step retrosynthetic methods on the widely used benchmark USPTO-50k \cite{liu2017retrosynthetic}.
% In Section \ref{sub:analysis}, we conduct further analysis on both datasets to delve deeper into the working behaviors of discrete latent variables.

\subsection{Evaluation Metrics}
We mainly use two metrics to evaluate the proposed \med, namely \textbf{top-$k$ accuracy} ($k\in\{1, 3, 5, 10\}$) and \textbf{coverage}.
As the most commonly used metric for retrosynthesis, top-$k$ accuracy is defined as the fraction of correctly predicted reactants with a ranking higher than $k$. 
We count an output sequence as correct only if it matches the ground truth SMILES exactly.
Apart from that, we use the coverage to assess whether \med~is able to model the multi-modal distribution. The coverage is defined as the average of the proportion of the correctly predicted ground truth reactions \cite{Kim_2021_Tied}.
Therefore, higher coverage indicates that the model is able to generate more accurate and diverse reactant candidates.

\subsection{Baselines} 
Following \cite{g2s}, we broadly divide single-step retrosynthetic baselines into two groups according to whether additional features or techniques are used.
\subsubsection{Retrosynthesis with additional features/techniques.} We classify template extraction, atom mapping, and data augmentation into additional features or techniques as each process is computationally expensive.
Template-based approaches \citep{Sacha_2021_MEGAN, retrox, dai2019retrosynthesis, Chen_Jung_2021_LocalRetro, Sun_2020_EBM} match the given prodcut with large-scale refined rules or reaction templates and apply them to the product to obtain a set of reactants in single-step retrosynthesis, which has been pointed out to be essentially a symbolic pattern recognition process \citep{segler2017neural}.
Graph-based approaches \citep{g2g, retrox, Seo_2021_GTA, Wang_2021_RetroPRIME, Chen_Jung_2021_LocalRetro, Somnath_2020_GRAPHRETRO, Sun_2020_EBM} are built upon atom mapping across chemical reactions, i.e., numbering atoms in a molecular graph to indicate which atom of the reactant(s) becomes which atom of the product(s) \citep{jaworski2019automatic}.
Aug. Transformer \citep{Tetko_2020_AT} and Chemformer \citep{Irwin_2021_Chemformer} explore fully data-driven single-step retrosynthesis by data augmentation, which increases top-$k$ prediction accuracy as well as training costs.

\subsubsection{Retrosynthesis without additional features/techniques.} 
The proposed \med~belongs to this category.
This line of research consists of AutoSynRoute \citep{Lin_2020_AutoSynRoute}, SCROP \citep{Zheng_2020_SCROP}, Latent Transformer \citep{Chen_2020_Latent}, GET \citep{Mao_2021_GET}, DMP fusion \citep{zhu2021dual}, Tied Transformer \citep{Kim_2021_Tied}, and Graph2SMILES \citep{g2s}.
All of them treat single-step retrosynthesis as a sequence-to-sequence task and are built upon the Transformer architecture.
% Latent Transformer \citep{Chen_2020_Latent} and Tied Transformer \citep{Kim_2021_Tied} are the only two models considering the retrosynthetic diversity.
% However, they both apply hard-EM algorithm, which is time-consuming and performs inferior than \med.
Note that Graph2SMILES \citep{g2s} is the closest to \med~in implementation, so we treat Graph2SMILES as the base model and compare it with \med~on extensive experiments on both public and homemade datasets.

\subsection{Empirical Methodology} 
Although CVAE is proved effective in modeling multi-modal distribution \cite{cvae}, \citet{DBLP:conf/conll/BowmanVVDJB16} point out that these types of models suffer from \textit{posterior collapse}, i.e., $\mathbf{z}$ and $\mathbf{x}$ become independent if we minimize the KL divergence directly.
To alleviate this problem, we use a simple yet effective approach, namely KL annealing \cite{DBLP:conf/conll/BowmanVVDJB16, DBLP:conf/nips/SonderbyRMSW16}, where a variable weight is added to the KL term in our objective function during training. Therefore, we aim to maximize the following objective function
\begin{align*}
    \Tilde{\mathcal{L}}_{\theta,\phi} (\mathbf{x},\mathbf{y})&=\log p_\theta (\mathbf{y}|\mathbf{x}, \mathbf{z})-\alpha D_{\text{KL}}\left(q_\phi(\mathbf{z}|\mathbf{x},\mathbf{y})\|p_\theta (\mathbf{z}|\mathbf{x}) \right),
\end{align*}
where the KL annealing coefficient $\alpha$ is gradually increased from 0 to 1 over training.

\subsection{Implementation details}
We implement our model using PyTorch and run it on a single NVIDIA GeForce 3090 GPU.
Both the encoder and decoder of our model are composed of 6 layers with 8 parallel attention heads.
The size of word embedding and atom representation is set to 256.
We instantiate the attention sublayers and the position-wise feed-forward network sublayers with
256 hidden units.
Following Graph2SMILES \cite{g2s}, we fix the filter size of Transformer to 2048.
We train \med~using Adam optimizer \cite{adam} with Noam learning rate scheduler \cite{vaswani2017attention}.
We apply a dropout rate of 0.3 and limit the number of epochs to 2000.
The layers of the recognition model and the prior model are both set to 1.
We use grid search to find the best latent sizes for different datasets among $\{10, 20, 30, 60, 120\}$.
Beam search is used to generate the output SMILES during inference with a beam size of 30.
To avoid over-tuning, we select the models based on their top-1 accuracy during the validation phase.

\section{Experimental Results}\label{sec:exp}
In this section, we conduct extensive experiments on two datasets to verify the effectiveness of \med. 
Our experiments are intended to answer the following three research questions.
\begin{itemize}[leftmargin=*]
\item \textbf{RQ1}: Can \med~model the multi-modal distribution over chemical reaction types as claimed?
\item \textbf{RQ2}: Does the proposed approach achieve competitive results on the benchmark dataset?
\item \textbf{RQ3}: What role do the discrete latent variables actually play in single-step retrosynthesis?
\end{itemize}

% \noindent\textbf{RQ I}: Can \med~alleviate the one-to-many problem as claimed?

% \noindent\textbf{RQ II}: Does the proposed approach achieve competitive results on the benchmark dataset?

% \noindent\textbf{RQ III}: What role do the discrete latent variables actually play in retrosynthesis?

\begin{figure*}[!ht]
\centering % <-- added
\includegraphics[width=0.85\linewidth]{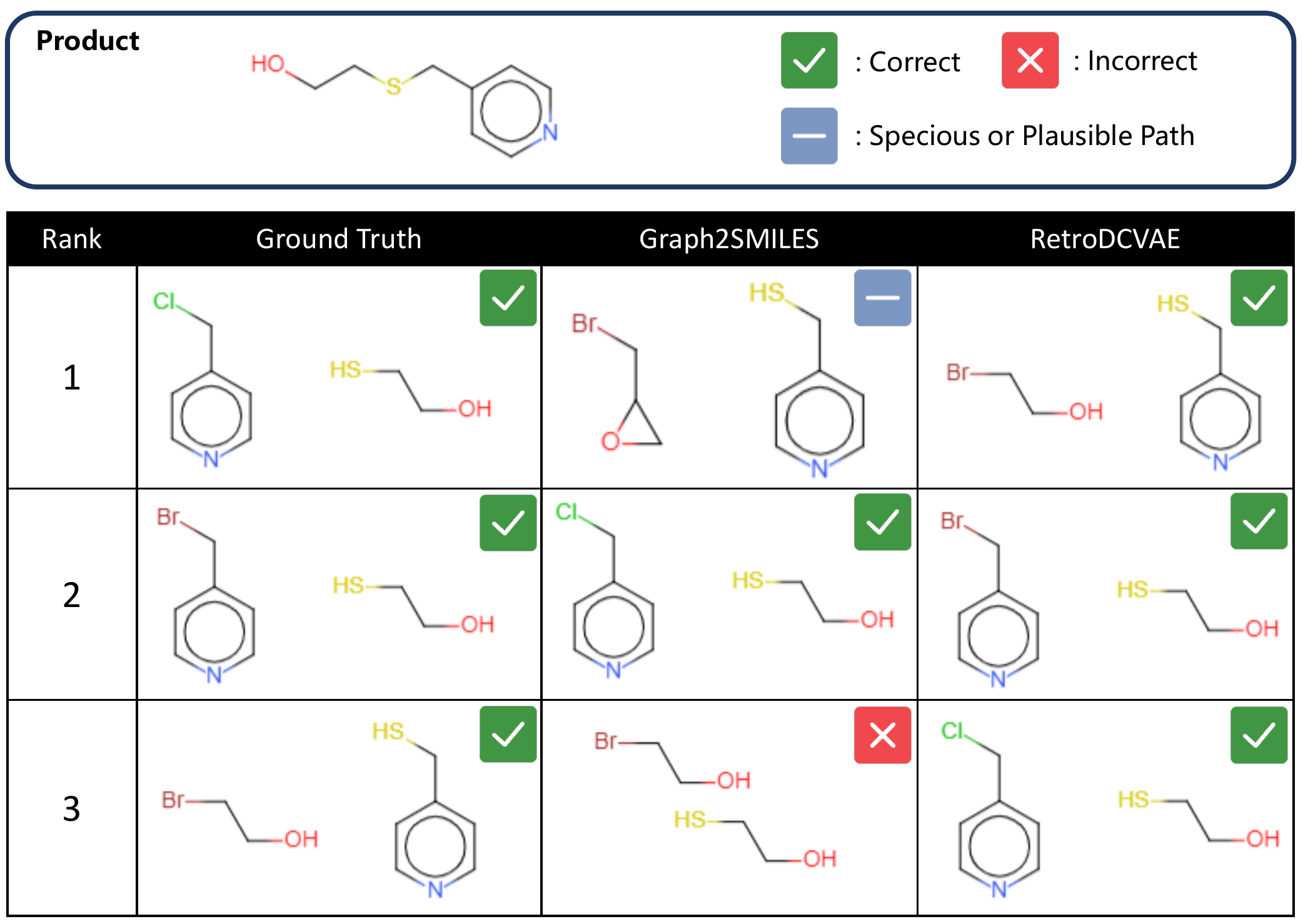}
\caption{Top-3 predictions of Graph2SMILES \cite{g2s} and \med.}
\label{fig:case}
\end{figure*}

\subsection{Evaluation on Homemade Dataset (RQ1)} \label{sub:uspto-diverse}
To assess whether \med~can model the multi-modal distribution as claimed, we  perform a training/validation/testing split as 8:1:1 over all 1-to-N products in USPTO-MIT \cite{jin2017predicting} to build USPTO-DIVERSE. 
The product of each reaction in USPTO-DIVERSE corresponds to multiple synthetic schemes, which poses a challenge for deterministic retrosynthetic approaches.
We summarize the comparison of {Graph2SMILES} \cite{g2s} and \med~in Table \ref{tab:uspto-diverse}, where the top-1 invalid rate means the proportion of grammatically invalid SMILES in top-1 predictions.
By leveraging probabilistic inference, \med~significantly reduces the invalid rate and consistently yields improvement in top-$k$ accuracy.

\begin{table}[ht]
\centering
    \caption{Model performance on USPTO-DIVERSE.}
    \begin{tabular}{@{}c ccccc @{}}
    \toprule
    \multirow{2}{*}{Model} & \multicolumn{4}{c}{Top-$k$ accuracy (\%)}&Top-1 invalid \\ 
    \cmidrule(l){2-5} 
    & 1 & 3 & 5 & 10 & rate (\%) \\ 
    \midrule
    Graph2SMILES \cite{g2s} & 32.2 & 45.2 & 47.9 & 52.1 & 8.20\\
    \med~& \textbf{33.4}& \textbf{49.8} & \textbf{54.1} & \textbf{57.1} & \textbf{4.77}\\
    \bottomrule
    \end{tabular}
    \label{tab:uspto-diverse}
    % \vspace{-3mm} 
\end{table}

We evaluate coverage of top-30 predictions on USPTO-DIVERSE as well.
Among 1105 products, 942 have two distinct ground truth reactions, 113 have three distinct ground truth reactions, and 11 have more than five ground truth reactions. 
Table \ref{tab:coverage} shows that the \med~covers on average 35.8\% of the actual reactions contained in the test set, which consistently improves the performance of Graph2SMILES \cite{g2s}.

\begin{table}[!ht]
\centering
    \caption{Coverage evaluation on USPTO-DIVERSE.}
    \begin{adjustbox}{max width=\linewidth}
    \begin{tabular}{c c c c }
    \toprule
    % \multirow{2}{*}{\#Reaction per product } &
    \#Reaction &\multirow{2}{*}{\#Product } & \multicolumn{2}{c}{Coverage (\%)} \\
     \cmidrule(l){3-4} 
    per product & & Graph2SMILES \cite{g2s} & \med\\ 
     \midrule
    2  & 942  & 33.9 & \textbf{36.6} \\
    3  & 113  & 30.4 & \textbf{33.6} \\
    4  & 35   & 23.6 & \textbf{28.6} \\
    5  & 4    & \textbf{10.0} & \textbf{10.0} \\
    6  & 4    & 8.3  & \textbf{16.7} \\
    7  & 3    & \textbf{23.8} & \textbf{23.8} \\
    8  & 2    & 6.3  & \textbf{18.9} \\
    9  & 1    & 11.1 & \textbf{22.2} \\
    % 10 & 1    & 0 & 0 \\0
    % 11 & 3    & 6.1 & 3.0   \\3.0
    13 & 1    & 7.7 & \textbf{15.4}  \\
    % 18 & 1    & 11.1 & 5.6  \\0
     \midrule
    average & &  32.9 & \textbf{35.8}  \\
    \bottomrule
    \end{tabular}
    \end{adjustbox}
    \label{tab:coverage}
    % \vspace{-3mm} 
\end{table}

To examine the quality of retrosynthetic prediction candidates, we compare the proposed \med~against Graph2SMILES \cite{g2s} in terms of top-3 predictions. Figure \ref{fig:case} shows the comparison results. For the same input product, both Graph2SMILES \cite{g2s} and \med~generate grammatically valid reactants. 
The proposed \med~correctly predicts three ground-truth paths, while Graph2SMILES \cite{g2s} generates chemically incorrect reactants and ranks a specious synthetic scheme higher than a correct one.
Therefore, the answer to \textbf{RQ1} is yes, i.e., \med~can alleviate the problem of the multi-modal distribution  as claimed.

\subsection{Evaluation on Public Dataset (RQ2)} \label{sub:uspto-50k}

\begin{table*}[h]
\caption{Retrosynthesis results on unlabeled USPTO-50k sorted by top-1 accuracy. \textit{Templ.}: reaction templates used; \textit{Map.}: atom-mapping required; \textit{Aug.}: output data augmentation used. Best results for each group of rows are highlighted in \textbf{bold}. The results of previous studies are taken from \cite{g2s}.}
\label{tab:uspto-50k}
\begin{center}
\begin{tabular}{l cccc ccc }
\toprule
\multirow{2}{*}{Methods} & \multicolumn{4}{c}{Top-$k$ accuracy (\%)} &
\multicolumn{3}{c}{Techniques used} \\
\cmidrule(lr){2-5} \cmidrule(lr){6-8}
&1  &3  &5  &10 
&\small \textit{Templ.} &\small \textit{Map.} &\small \textit{Aug.}\\
\midrule
AutoSynRoute \citep{Lin_2020_AutoSynRoute}
&43.1   &64.6   &71.8   &\textbf{78.7}   &\xmark &\xmark &\xmark \\
SCROP \citep{Zheng_2020_SCROP}
&43.7   &60.0   &65.2   &68.7   &\xmark &\xmark &\xmark \\
Latent Transformer \citep{Chen_2020_Latent}
&44.8   &64.9   &72.4   &78.4   &\xmark &\xmark &\xmark \\
GET \citep{Mao_2021_GET}
&44.9   &58.8   &62.4   &65.9   &\xmark &\xmark &\xmark \\
DMP fusion \citep{zhu2021dual}
&46.1   &65.2   &70.4   &74.3   &\xmark &\xmark &\xmark \\
Tied Transformer \citep{Kim_2021_Tied}
&47.1   &67.2   &\textbf{73.5}   &78.5   &\xmark &\xmark &\xmark \\
Graph2SMILES  \citep{g2s} %(D-GCN)
&52.9   &66.5   &70.0   &72.9   &\xmark &\xmark &\xmark \\
\med~(\textit{ours})
&\textbf{53.1}   &\textbf{68.1}   &71.6   &74.3   &\xmark &\xmark &\xmark \\
\midrule
MEGAN \citep{Sacha_2021_MEGAN}
&48.1   &70.7   &78.4   &86.1   &\xmark &\cmark &\xmark \\
G2Gs \citep{g2g}
&48.9   &67.6   &72.5   &75.5   &\xmark &\cmark &\xmark \\
RetroXpert \citep{retrox}
&50.4   &61.1   &62.3   &63.4   &\cmark &\cmark &\cmark \\
GTA \citep{Seo_2021_GTA}
&51.1   &67.6   &74.8   &81.6   &\xmark &\cmark &\cmark \\
RetroPrime \citep{Wang_2021_RetroPRIME}
&51.4   &70.8   &74.0   &76.1   &\xmark &\cmark &\cmark \\
GLN \citep{dai2019retrosynthesis}
&52.5   &69.0   &75.6   &83.7   &\cmark &\cmark &\xmark \\
Aug. Transformer \citep{Tetko_2020_AT}
&53.2   &-      &80.5   &85.2   &\xmark &\xmark &\cmark \\
% EBM  (Dual-TF) \citep{Sun_2020_EBM}
% &53.3   &69.7   &73.0   &75.0   &\xmark &\xmark &\cmark \tabularnewline
LocalRetro \citep{Chen_Jung_2021_LocalRetro}
&53.4   &\textbf{77.5}   &\textbf{85.9}   &\textbf{92.4}   &\cmark &\cmark &\xmark \\
GraphRetro \citep{Somnath_2020_GRAPHRETRO}
&53.7   &68.3   &72.2   &75.5   &\xmark &\cmark &\xmark \\
Chemformer \citep{Irwin_2021_Chemformer}
&54.3   &-      &62.3   &63.0   &\xmark &\xmark &\cmark \\
EBM (Dual-TB) \citep{Sun_2020_EBM}
&\textbf{55.2}   &74.6   &80.5   &86.9   &\cmark &\cmark &\cmark \\
\bottomrule
\end{tabular}
\end{center}
\end{table*}

\begin{figure*}[!ht]
  \includegraphics[width=\linewidth]{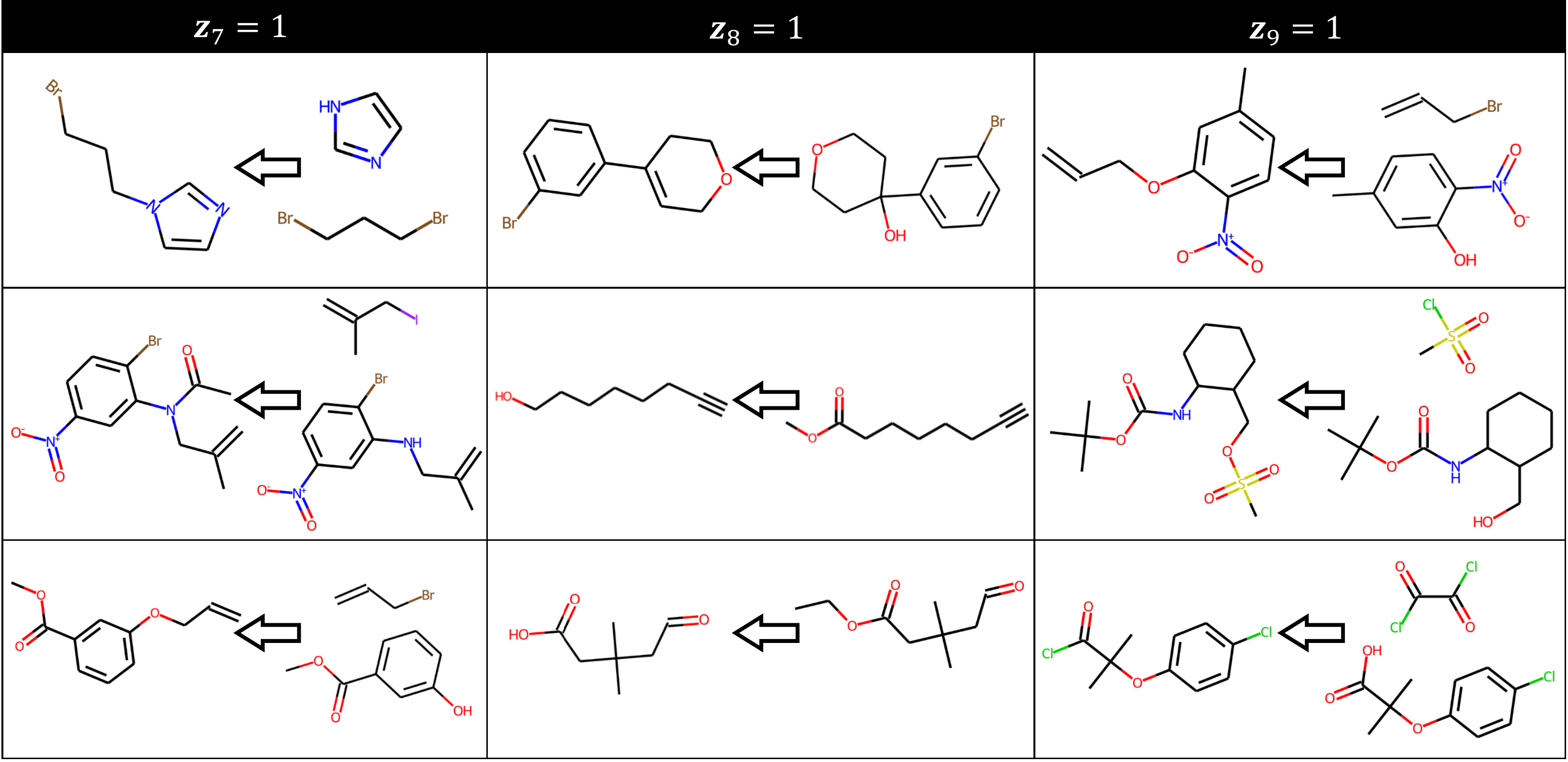}
\caption{Reactions in USPTO-DIVERSE that corresponding to different $\mathbf{z}$. As $\mathbf{z}$ is a discrete latent variable, $\mathbf{z}_i=1$ indicates that $\mathbf{z}$ is the one-hot vector $\mathbf{e}_i$. \med~reads the product SMILES then generates $\mathbf{z}$ and reactants.}
\label{fig:latent}
\vspace{-3.5mm}
\end{figure*}

Table \ref{tab:uspto-50k} summarizes the top-$k$ accuracy evaluated on the benchmark dataset USPTO-50k. 
We divide existing single-step retrosynthetic approaches into two groups.
The best top-$k$ accuracy of each group is highlighted with boldface.

\begin{figure*}[!ht]
  \includegraphics[width=\linewidth]{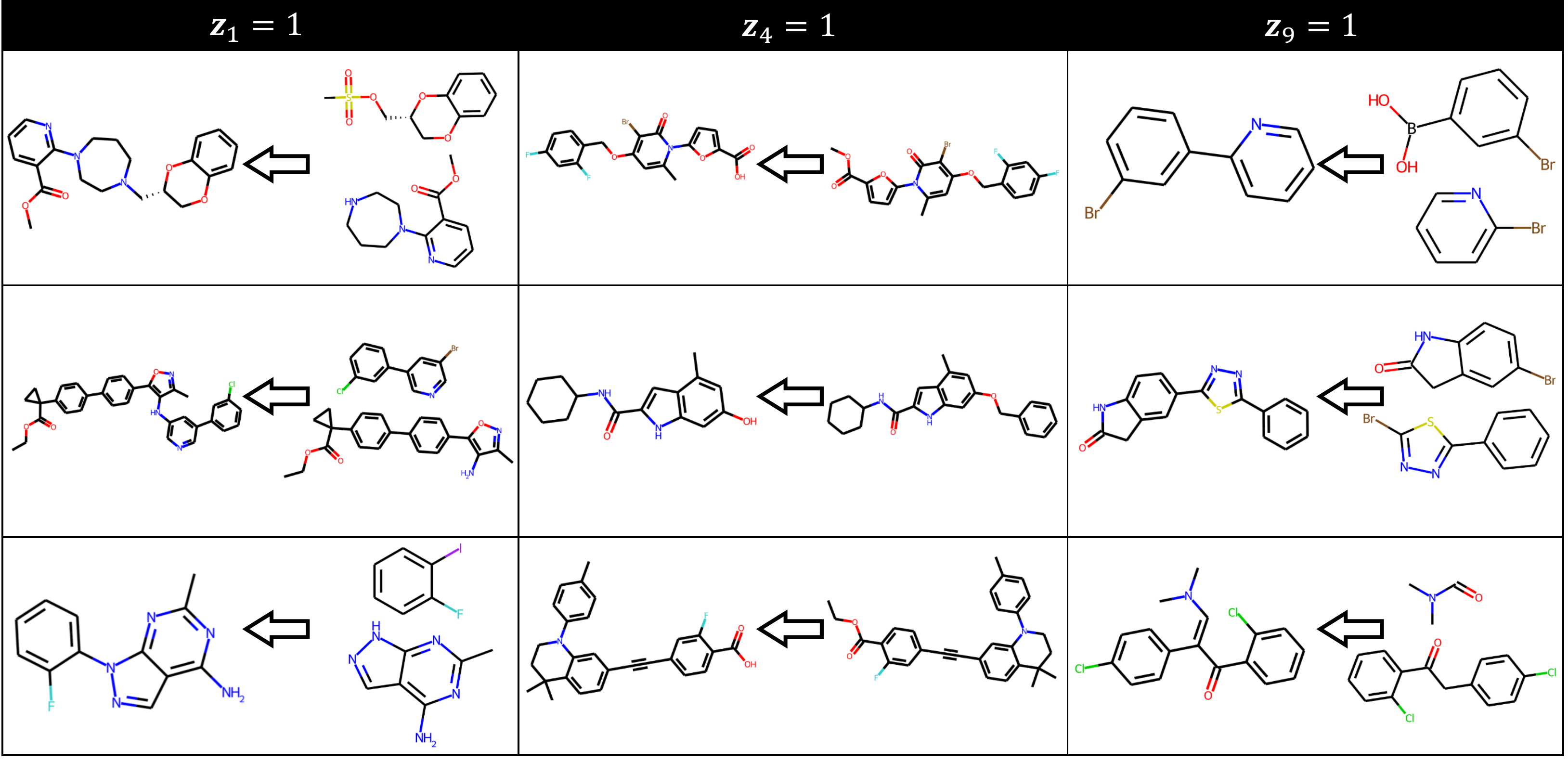}
\caption{Reactions in USPTO-50k that corresponding to different latent variables. Here $\mathbf{z}_i=1$ indicates that $\mathbf{z}$ is the one-hot vector $\mathbf{e}_i$. \med~reads the product SMILES then generates $\mathbf{z}$ and reactants.}
\label{fig:latent_50k}
\vspace{-3.5mm}
\end{figure*}

Overall, \med~achieves competitive results for single-step retrosynthesis and consistently outperforms Graph2SMILES \cite{g2s} on top-$k$ accuracy. 
The improvement lies in that \med~introduces stochastic inference to structured output prediction, which lowers the risk of over-fitting and reduces the probability of invalid SMILES string output. 

Following \citep{g2s}, we divide single-step retrosynthetic methods into two groups according to whether the additional techniques such as templates, atom mapping, and output-side data augmentation are used.
Table \ref{tab:uspto-50k} shows that \med~achieves the best top-1 and top-3 accuracies across methods that do not use reaction templates, atom mapping, or output SMILES augmentation.
That is, \med~achieves state-of-the-art performance while exempting the need of domain knowledge.

Compared to the second group in Table \ref{tab:uspto-50k}, \med~further achieves comparable results against the methods that use additional features or techniques. Specifically, \med~even outperforms several template-based baselines \cite{retrox, dai2019retrosynthesis} in terms of the top-1 accuracy.
Therefore, the answer to \textbf{RQ2} is yes.
We conclude that \med~is able to achieve competitive results on the benchmark dataset USPTO-50k.

\subsection{Further Analysis on Latent Variables (RQ3)} \label{sub:analysis}
As mentioned in Section \ref{sec:med}, we use the Gumbel-Softmax distribution to approximate categorical samples and discretize the latent variable $\mathbf{z}$ with Straight-Through Gumbel Estimator \cite{gumbel}.
To gain insights into the working behaviors of \med, we conduct qualitative analysis on the discrete latent variable $\mathbf{z}$ on both public and homemade datasets.

\med~reads an arbitrary product sequence and encodes it using a topology-aware encoder, then the latent variable $\mathbf{z}$ is sampled according to the product embeddings.
Conditioned on $\mathbf{z}$, \med~generates possible reactants using its variational decoder. To figure out what role the latent variables play, we collect the reactions that associate with the same one-hot vector $\mathbf{z}$ and present examples in following sections.
% Figure \ref{fig:latent}.

\subsubsection{Qualitative Examples on USPTO-DIVERSE} \label{sub:qua-diverse}
As depicted in Figure \ref{fig:latent}, reactions corresponding to the same $\mathbf{z}$ share many common characteristics.
We follow the chemical reaction classification criteria adopted by \citet{schneider2016s}.
The reactions in the first column can be classified into functional group interconversion (FGI). Moreover, the reactants of the three reactions all contain bromine (Br) atoms.
Reactions corresponding to $\mathbf{z}=\mathbf{e}_8$ in the second column are associated with the formation or disappearance of the C-O bond. In addition, each reaction in the second column only has one reactant.
As for the third column, the reactions corresponding to $\mathbf{z}=\mathbf{e}_9$ can be categorized into functional group addition (FGA).
% Besides the analysis on USPTO-DIVERSE, please refer to Appendix \ref{sec:app_A} for examples of the benchmark dataset USPTO-50k.

Therefore, the answer to \textbf{RQ3} becomes clear.
In general, we can view the discrete latent variables as reaction indicators that guide the generation of distinct reactants.
The discrete latent variables reduce grammatically invalid rates and increase reaction diversity for single-step retrosynthesis.
% Moreover, the proposed discrete latent variables may have the potential to automatically classify unlabeled reactions in public patents granted by USPTO, reducing the need for experience and expertise.

\subsubsection{Qualitative Examples on USPTO-50k}
\label{sec:app_A}
In addition to the qualitative examples on USPTO-DIVERSE, we analyse the discrete latent variable $\mathbf{z}$ on the benchmark dataset USPTO-50k \cite{liu2017retrosynthetic} as well.
Figure \ref{fig:latent_50k} presents some reaction examples that correspond to different discrete latent variables.
Following the reaction classification criteria in \cite{schneider2016s}, we summarize the reactions in each column as follows.

The reactions corresponding to $\mathbf{z}=\mathbf{e}_1$ in the first column are associated with the formation of the C-C bond.
And the reactions in the second column belong to the deprotection reaction, where a protecting group is modified by converting it into a functional group.
Finally, reactions in the third column are all heteroatom alkylation and arylation reactions.
Therefore, the conclusion in Section \ref{sub:qua-diverse} still holds.
The discrete latent variable $\mathbf{z}$ works as a reaction indicator that guides the generation of distinct sets of reactants.
Moreover, the proposed discrete latent variables may have the potential to automatically classify unlabeled reactions in granted patents, reducing the need for experience and expertise.

\section{Discussion} \label{sec:dis}
In this work, we present \med, a novel single-step retrosynthetic approach inspired by CVAE.
To delve deeper into the property of \med, we discuss the advantages and limitation of \med~in Section \ref{sub:advantage} and Section \ref{sub:limit}, respectively.
\subsection{Advantages of \med}\label{sub:advantage}
The proposed \med~aims to model the multi-modal distribution over reaction types via discrete latent variables.
Compared with state-of-the-art single-step retrosynthetic models \citep{Chen_Jung_2021_LocalRetro, Somnath_2020_GRAPHRETRO, Irwin_2021_Chemformer, Sun_2020_EBM}, \med~reduces the need for reaction templates, atom mapping and data augmentation, while maintaining comparable retrosynthetic performance.
Compared with existing sequence-based approaches, \med~achieves the best top-$k$ accuracy $(k=1, 3)$ for single-step retrosynthesis, while being among the most efficient to train and predict.

Apart from its effectiveness, another significant advantage of \med~is its interpretability.
Most existing template-free methods simply output reactant candidates without any explanation.
Different from them, \med~associates each set of predicted reactants with a discrete latent variable, which we interpret as the corresponding reaction type.
The interpretability of \med~is expected to have an immediate and strong impact on computer-aided retrosynthetic tasks.

\subsection{Limitation of \med}\label{sub:limit}
% Although \med~has achieved competitive retrosynthetic performance, it can be further improved and extended.
One limitation of this work is that \med~does not consider catalysts or reagents when generating reactant candidates for the desired product, which is a common fault of existing computer-aided retrosynthetic approaches.
In fact, a chemical reaction is related not only to reactants and products, but also to reaction conditions including catalysts, temperature, and reagent concentration.
The reaction conditions indicate the difficulty of synthesizing the product to some extent, which is important for retrosynthesis planning.
Although it is a pioneer to model reaction types, \med~lacks the ability to predict detailed reaction conditions, which can be further improved and extended.
\section{Related Work}
Our work is closely related to single-step retrosynthetic approaches and CVAE based natural language processing (NLP).
In this part, we first review single-step retrosynthetic methods in Section \ref{sec:one-step},
then briefly introduce CVAE and its variants in Section \ref{sec:cvaes}.

\subsection{Single-step Retrosynthesis} \label{sec:one-step}
An increasing number of machine learning approaches have emerged to assist in designing synthetic schemes for a target product. 
We broadly categorize them into template-based and template-free.
\subsubsection{Template-based retrosynthesis.} Relying on domain knowledge, template-based methods \cite{coley2017computer,dai2019retrosynthesis,segler2017neural,retrox,Chen_Jung_2021_LocalRetro,Sun_2020_EBM} match the given products with large-scale chemical rules or reaction templates. 
These rules can be hand-encoded by human experts or automatically extracted from data. This procedure is challenging because encoding which adjacent functional groups influence the outcome of a reaction requires an understanding of the underlying mechanisms.
Despite their interpretability, template-based methods are criticized for poor generalization to new and rare reactions \cite{Somnath_2020_GRAPHRETRO,Lin_2020_AutoSynRoute}, which entails a paradigm shift to template-free approaches.
\subsubsection{Template-free retrosynthesis.}
According to the leading data format during training, template-free approaches fall into two categories, namely graph edit-based and translation-based.
The first category views retrosynthesis as graph transformations.
Specifically, \citep{g2g, retrox, Somnath_2020_GRAPHRETRO, Wang_2021_RetroPRIME} first identify reaction centers, then perform graph or sequence recovery.
Translation-based methods formulate the product-to-reactant process of single-step retrosynthesis as SMILES-to-SMILES translation, typically with sequence models such as Transformer \cite{Lin_2020_AutoSynRoute, yang2019molecular, duan2020retrosynthesis, Tetko_2020_AT}.
Some variants of the second category perform pretraining, rerank the predictions, or leverage graph topology to enhance performance \cite{Irwin_2021_Chemformer, zhu2021dual, Sun_2020_EBM, g2s}.
% 区分我们方法和已有方法的不同？

\subsection{Conditional Generative Models in NLP} \label{sec:cvaes}
Formulating single-step retrosynthesis as machine translation, we draw inspiration from conditional generative models in NLP.
\citet{cvae} propose CVAE to perform probabilistic inference and make diverse predictions, where the Gaussian latent variables are used to model complex structured output representations.
\citet{cvae4nmt} introduce CVAE to neural machine translation for conditional text generation, 
while \citet{acl17} and \citet{aaai17} apply continuous latent variables to diverse dialogue generation.
In addition, MojiTalk \cite{mojitalk} leverages CVAE variants to explicitly control the emotion and sentiment of generated text.
\citet{aaai20} propose a multi-pass hierarchical conditional variational autoencoder to improve automatic storytelling.
However, the aforementioned methods only consider continuous latent variables, which is unfit for retrosynthesis given the discrete nature of chemical reactions.
Unlike them, we use discrete latent variables to guide reactant generation and leverage Gumbel-Softmax approximation to support backpropagation, leading to the novelty of this work.

\section{Conclusion}
In this paper, we present \med, a template-free retrosynthesizer that is able to generate diverse reactant candidates via discrete conditional variational autoencoders. 
% \med~incorporates conditional variational autoencoders into the retrosynthesis and uses discrete latent variables to guide the reactant generation.
Extensive experiments on both public and homemade datasets demonstrate that \med~consistently outperforms template-free baselines on single-step retrosynthesis.
One limitation of \med~is that we do not take into account factors such as catalysts and reagent concentration during the synthetic process, which is a common fault of existing computer-aided retrosynthetic approaches.
In future work, we aim to consider detailed reaction conditions and extend \med~to multi-step retrosynthesis, i.e., the retrosynthesis planning task.
We anticipate that \med~would exert potential impacts on modern drug discovery, particularly in accelerating the development of treatments and drugs for COVID-19.

\newpage

\bibliographystyle{ACM-Reference-Format}
\bibliography{custom}


\begin{thebibliography}{55}

%%% the Web address.

\ifx \showCODEN    \undefined \def \showCODEN     #1{\unskip}     \fi
\ifx \showDOI      \undefined \def \showDOI       #1{#1}\fi
\ifx \showISBNx    \undefined \def \showISBNx     #1{\unskip}     \fi
\ifx \showISBNxiii \undefined \def \showISBNxiii  #1{\unskip}     \fi
\ifx \showISSN     \undefined \def \showISSN      #1{\unskip}     \fi
\ifx \showLCCN     \undefined \def \showLCCN      #1{\unskip}     \fi
\ifx \shownote     \undefined \def \shownote      #1{#1}          \fi
\ifx \showarticletitle \undefined \def \showarticletitle #1{#1}   \fi
\ifx \showURL      \undefined \def \showURL       {\relax}        \fi
% invisible to TeX
\providecommand\bibfield[2]{#2}
\providecommand\bibinfo[2]{#2}
\providecommand\natexlab[1]{#1}
\providecommand\showeprint[2][]{arXiv:#2}

\bibitem[Bowman et~al\mbox{.}(2016)]%
        {DBLP:conf/conll/BowmanVVDJB16}
\bibfield{author}{\bibinfo{person}{Samuel~R. Bowman}, \bibinfo{person}{Luke
  Vilnis}, \bibinfo{person}{Oriol Vinyals}, \bibinfo{person}{Andrew~M. Dai},
  \bibinfo{person}{Rafal J{\'{o}}zefowicz}, {and} \bibinfo{person}{Samy
  Bengio}.} \bibinfo{year}{2016}\natexlab{}.
\newblock \showarticletitle{Generating Sentences from a Continuous Space}. In
  \bibinfo{booktitle}{\emph{Proc. of CoNLL}}.
\newblock


\bibitem[Chen et~al\mbox{.}(2019)]%
        {Chen_2020_Latent}
\bibfield{author}{\bibinfo{person}{Benson Chen}, \bibinfo{person}{Tianxiao
  Shen}, \bibinfo{person}{Tommi~S Jaakkola}, {and} \bibinfo{person}{Regina
  Barzilay}.} \bibinfo{year}{2019}\natexlab{}.
\newblock \showarticletitle{Learning to make generalizable and diverse
  predictions for retrosynthesis}.
\newblock \bibinfo{journal}{\emph{arXiv preprint arXiv:1910.09688}}
  (\bibinfo{year}{2019}).
\newblock


\bibitem[Chen and Jung(2021)]%
        {Chen_Jung_2021_LocalRetro}
\bibfield{author}{\bibinfo{person}{Shuan Chen} {and} \bibinfo{person}{Yousung
  Jung}.} \bibinfo{year}{2021}\natexlab{}.
\newblock \showarticletitle{Deep Retrosynthetic Reaction Prediction using Local
  Reactivity and Global Attention}.
\newblock \bibinfo{journal}{\emph{JACS Au}} (\bibinfo{year}{2021}).
\newblock


\bibitem[Cho et~al\mbox{.}(2014)]%
        {gru}
\bibfield{author}{\bibinfo{person}{Kyunghyun Cho}, \bibinfo{person}{Bart van
  Merrienboer}, \bibinfo{person}{Dzmitry Bahdanau}, {and}
  \bibinfo{person}{Yoshua Bengio}.} \bibinfo{year}{2014}\natexlab{}.
\newblock \showarticletitle{On the Properties of Neural Machine Translation:
  Encoder-Decoder Approaches}. In \bibinfo{booktitle}{\emph{Proceedings of
  SSST@EMNLP 2014}}.
\newblock


\bibitem[Coley et~al\mbox{.}(2017)]%
        {coley2017computer}
\bibfield{author}{\bibinfo{person}{Connor~W Coley}, \bibinfo{person}{Luke
  Rogers}, \bibinfo{person}{William~H Green}, {and} \bibinfo{person}{Klavs~F
  Jensen}.} \bibinfo{year}{2017}\natexlab{}.
\newblock \showarticletitle{Computer-assisted retrosynthesis based on molecular
  similarity}.
\newblock \bibinfo{journal}{\emph{ACS central science}} (\bibinfo{year}{2017}).
\newblock


\bibitem[Corey and Wipke(1969)]%
        {corey1969computer}
\bibfield{author}{\bibinfo{person}{Elias~James Corey} {and}
  \bibinfo{person}{W~Todd Wipke}.} \bibinfo{year}{1969}\natexlab{}.
\newblock \showarticletitle{Computer-assisted design of complex organic
  syntheses}.
\newblock \bibinfo{journal}{\emph{Science}} (\bibinfo{year}{1969}).
\newblock


\bibitem[Dai et~al\mbox{.}(2019a)]%
        {dai2019retrosynthesis}
\bibfield{author}{\bibinfo{person}{Hanjun Dai}, \bibinfo{person}{Chengtao Li},
  \bibinfo{person}{Connor Coley}, \bibinfo{person}{Bo Dai}, {and}
  \bibinfo{person}{Le Song}.} \bibinfo{year}{2019}\natexlab{a}.
\newblock \showarticletitle{Retrosynthesis Prediction with Conditional Graph
  Logic Network}. In \bibinfo{booktitle}{\emph{Proc. of NeurIPS}}.
\newblock


\bibitem[Dai et~al\mbox{.}(2019b)]%
        {transformer-xl}
\bibfield{author}{\bibinfo{person}{Zihang Dai}, \bibinfo{person}{Zhilin Yang},
  \bibinfo{person}{Yiming Yang}, \bibinfo{person}{Jaime~G. Carbonell},
  \bibinfo{person}{Quoc~Viet Le}, {and} \bibinfo{person}{Ruslan
  Salakhutdinov}.} \bibinfo{year}{2019}\natexlab{b}.
\newblock \showarticletitle{Transformer-{XL}: Attentive Language Models beyond
  a Fixed-Length Context}. In \bibinfo{booktitle}{\emph{Proc. of ACL}}.
\newblock


\bibitem[Duan et~al\mbox{.}(2020)]%
        {duan2020retrosynthesis}
\bibfield{author}{\bibinfo{person}{Hongliang Duan}, \bibinfo{person}{Ling
  Wang}, \bibinfo{person}{Chengyun Zhang}, \bibinfo{person}{Lin Guo}, {and}
  \bibinfo{person}{Jianjun Li}.} \bibinfo{year}{2020}\natexlab{}.
\newblock \showarticletitle{Retrosynthesis with attention-based NMT model and
  chemical analysis of “wrong” predictions}.
\newblock \bibinfo{journal}{\emph{RSC advances}} (\bibinfo{year}{2020}).
\newblock


\bibitem[Gershman and Goodman(2014)]%
        {avi}
\bibfield{author}{\bibinfo{person}{Samuel Gershman} {and}
  \bibinfo{person}{Noah~D. Goodman}.} \bibinfo{year}{2014}\natexlab{}.
\newblock \showarticletitle{Amortized Inference in Probabilistic Reasoning}. In
  \bibinfo{booktitle}{\emph{Proc. of CogSci}}.
\newblock


\bibitem[Gumbel(1954)]%
        {gumbel1954statistical}
\bibfield{author}{\bibinfo{person}{Emil~Julius Gumbel}.}
  \bibinfo{year}{1954}\natexlab{}.
\newblock \bibinfo{booktitle}{\emph{Statistical theory of extreme values and
  some practical applications: a series of lectures}}.
\newblock \bibinfo{publisher}{US Government Printing Office}.
\newblock


\bibitem[Hendrycks and Gimpel(2016)]%
        {gelu}
\bibfield{author}{\bibinfo{person}{Dan Hendrycks} {and} \bibinfo{person}{Kevin
  Gimpel}.} \bibinfo{year}{2016}\natexlab{}.
\newblock \showarticletitle{Bridging Nonlinearities and Stochastic Regularizers
  with Gaussian Error Linear Units}.
\newblock \bibinfo{journal}{\emph{arXiv preprint arXiv:1606.08415}}
  (\bibinfo{year}{2016}).
\newblock


\bibitem[Hu et~al\mbox{.}(2020)]%
        {gnn-ptm}
\bibfield{author}{\bibinfo{person}{Weihua Hu}, \bibinfo{person}{Bowen Liu},
  \bibinfo{person}{Joseph Gomes}, \bibinfo{person}{Marinka Zitnik},
  \bibinfo{person}{Percy Liang}, \bibinfo{person}{Vijay~S. Pande}, {and}
  \bibinfo{person}{Jure Leskovec}.} \bibinfo{year}{2020}\natexlab{}.
\newblock \showarticletitle{Strategies for Pre-training Graph Neural Networks}.
  In \bibinfo{booktitle}{\emph{Proc. of ICLR}}.
\newblock


\bibitem[Irwin et~al\mbox{.}(2021)]%
        {Irwin_2021_Chemformer}
\bibfield{author}{\bibinfo{person}{R. Irwin}, \bibinfo{person}{S. Dimitriadis},
  \bibinfo{person}{J. He}, {and} \bibinfo{person}{E. Bjerrum}.}
  \bibinfo{year}{2021}\natexlab{}.
\newblock \bibinfo{title}{Chemformer: A Pre-Trained Transformer for
  Computational Chemistry}.
\newblock
\newblock


\bibitem[Jang et~al\mbox{.}(2017)]%
        {gumbel}
\bibfield{author}{\bibinfo{person}{Eric Jang}, \bibinfo{person}{Shixiang Gu},
  {and} \bibinfo{person}{Ben Poole}.} \bibinfo{year}{2017}\natexlab{}.
\newblock \showarticletitle{Categorical Reparameterization with
  Gumbel-Softmax}. In \bibinfo{booktitle}{\emph{Proc. of ICLR}}.
\newblock


\bibitem[Jaworski et~al\mbox{.}(2019)]%
        {jaworski2019automatic}
\bibfield{author}{\bibinfo{person}{Wojciech Jaworski}, \bibinfo{person}{Sara
  Szymku{\'c}}, \bibinfo{person}{Barbara Mikulak-Klucznik},
  \bibinfo{person}{Krzysztof Piecuch}, \bibinfo{person}{Tomasz Klucznik},
  \bibinfo{person}{Micha{\l} Ka{\'z}mierowski}, \bibinfo{person}{Jan
  Rydzewski}, \bibinfo{person}{Anna Gambin}, \bibinfo{person}{Bartosz~A
  Grzybowski}, {et~al\mbox{.}}} \bibinfo{year}{2019}\natexlab{}.
\newblock \showarticletitle{Automatic mapping of atoms across both simple and
  complex chemical reactions}.
\newblock \bibinfo{journal}{\emph{Nat. Comm.}} (\bibinfo{year}{2019}).
\newblock


\bibitem[Jin et~al\mbox{.}(2018)]%
        {jin2018junction}
\bibfield{author}{\bibinfo{person}{Wengong Jin}, \bibinfo{person}{Regina
  Barzilay}, {and} \bibinfo{person}{Tommi Jaakkola}.}
  \bibinfo{year}{2018}\natexlab{}.
\newblock \showarticletitle{Junction tree variational autoencoder for molecular
  graph generation}. In \bibinfo{booktitle}{\emph{Proc. of ICML}}.
\newblock


\bibitem[Jin et~al\mbox{.}(2017)]%
        {jin2017predicting}
\bibfield{author}{\bibinfo{person}{Wengong Jin}, \bibinfo{person}{Connor~W.
  Coley}, \bibinfo{person}{Regina Barzilay}, {and} \bibinfo{person}{Tommi~S.
  Jaakkola}.} \bibinfo{year}{2017}\natexlab{}.
\newblock \showarticletitle{Predicting Organic Reaction Outcomes with
  Weisfeiler-Lehman Network}. In \bibinfo{booktitle}{\emph{Proc. of NeurIPS}}.
\newblock


\bibitem[Kim et~al\mbox{.}(2021)]%
        {Kim_2021_Tied}
\bibfield{author}{\bibinfo{person}{Eunji Kim}, \bibinfo{person}{Dongseon Lee},
  \bibinfo{person}{Youngchun Kwon}, \bibinfo{person}{Min~Sik Park}, {and}
  \bibinfo{person}{Youn-Suk Choi}.} \bibinfo{year}{2021}\natexlab{}.
\newblock \showarticletitle{Valid, Plausible, and Diverse Retrosynthesis Using
  Tied Two-Way Transformers with Latent Variables}.
\newblock \bibinfo{journal}{\emph{J. Chem. Inf. Model.}}
  (\bibinfo{year}{2021}).
\newblock


\bibitem[Kingma and Ba(2015)]%
        {adam}
\bibfield{author}{\bibinfo{person}{Diederick~P Kingma} {and}
  \bibinfo{person}{Jimmy Ba}.} \bibinfo{year}{2015}\natexlab{}.
\newblock \showarticletitle{Adam: A method for stochastic optimization}. In
  \bibinfo{booktitle}{\emph{Proc. of ICLR}}.
\newblock


\bibitem[Kingma and Welling(2014)]%
        {vae}
\bibfield{author}{\bibinfo{person}{Diederik~P. Kingma} {and}
  \bibinfo{person}{Max Welling}.} \bibinfo{year}{2014}\natexlab{}.
\newblock \showarticletitle{Auto-Encoding Variational Bayes}. In
  \bibinfo{booktitle}{\emph{Proc. of ICLR}}.
\newblock


\bibitem[Lin et~al\mbox{.}(2020)]%
        {Lin_2020_AutoSynRoute}
\bibfield{author}{\bibinfo{person}{Kangjie Lin}, \bibinfo{person}{Youjun Xu},
  \bibinfo{person}{Jianfeng Pei}, {and} \bibinfo{person}{Luhua Lai}.}
  \bibinfo{year}{2020}\natexlab{}.
\newblock \showarticletitle{Automatic retrosynthetic route planning using
  template-free models}.
\newblock \bibinfo{journal}{\emph{Chem. Sci.}} (\bibinfo{year}{2020}).
\newblock


\bibitem[Liu et~al\mbox{.}(2017)]%
        {liu2017retrosynthetic}
\bibfield{author}{\bibinfo{person}{Bowen Liu}, \bibinfo{person}{Bharath
  Ramsundar}, \bibinfo{person}{Prasad Kawthekar}, \bibinfo{person}{Jade Shi},
  \bibinfo{person}{Joseph Gomes}, \bibinfo{person}{Quang Luu~Nguyen},
  \bibinfo{person}{Stephen Ho}, \bibinfo{person}{Jack Sloane},
  \bibinfo{person}{Paul Wender}, {and} \bibinfo{person}{Vijay Pande}.}
  \bibinfo{year}{2017}\natexlab{}.
\newblock \showarticletitle{Retrosynthetic reaction prediction using neural
  sequence-to-sequence models}.
\newblock \bibinfo{journal}{\emph{ACS central science}} (\bibinfo{year}{2017}).
\newblock


\bibitem[Liu et~al\mbox{.}(2020)]%
        {retrognn}
\bibfield{author}{\bibinfo{person}{Cheng-Hao Liu}, \bibinfo{person}{Maksym
  Korablyov}, \bibinfo{person}{Stanis{\l}aw Jastrz{\k{e}}bski},
  \bibinfo{person}{Pawe{\l} W{\l}odarczyk-Pruszy{\'n}ski},
  \bibinfo{person}{Yoshua Bengio}, {and} \bibinfo{person}{Marwin~HS Segler}.}
  \bibinfo{year}{2020}\natexlab{}.
\newblock \showarticletitle{RetroGNN: Approximating Retrosynthesis by Graph
  Neural Networks for De Novo Drug Design}.
\newblock \bibinfo{journal}{\emph{arXiv preprint arXiv:2011.13042}}
  (\bibinfo{year}{2020}).
\newblock


\bibitem[Maddison et~al\mbox{.}(2014)]%
        {maddison2014sampling}
\bibfield{author}{\bibinfo{person}{Chris~J Maddison}, \bibinfo{person}{Daniel
  Tarlow}, {and} \bibinfo{person}{Tom Minka}.} \bibinfo{year}{2014}\natexlab{}.
\newblock \showarticletitle{A* Sampling}. In \bibinfo{booktitle}{\emph{Proc. of
  NeurIPS}}.
\newblock


\bibitem[Mao et~al\mbox{.}(2021)]%
        {Mao_2021_GET}
\bibfield{author}{\bibinfo{person}{Kelong Mao}, \bibinfo{person}{Xi Xiao},
  \bibinfo{person}{Tingyang Xu}, \bibinfo{person}{Yu Rong},
  \bibinfo{person}{Junzhou Huang}, {and} \bibinfo{person}{Peilin Zhao}.}
  \bibinfo{year}{2021}\natexlab{}.
\newblock \showarticletitle{Molecular graph enhanced transformer for
  retrosynthesis prediction}.
\newblock \bibinfo{journal}{\emph{Neurocomputing}} (\bibinfo{year}{2021}).
\newblock


\bibitem[Pagnoni et~al\mbox{.}(2018)]%
        {cvae4nmt}
\bibfield{author}{\bibinfo{person}{Artidoro Pagnoni}, \bibinfo{person}{Kevin
  Liu}, {and} \bibinfo{person}{Shangyan Li}.} \bibinfo{year}{2018}\natexlab{}.
\newblock \showarticletitle{Conditional variational autoencoder for neural
  machine translation}.
\newblock \bibinfo{journal}{\emph{arXiv preprint arXiv:1812.04405}}
  (\bibinfo{year}{2018}).
\newblock


\bibitem[Rezende et~al\mbox{.}(2014)]%
        {DBLP:conf/icml/RezendeMW14}
\bibfield{author}{\bibinfo{person}{Danilo~Jimenez Rezende},
  \bibinfo{person}{Shakir Mohamed}, {and} \bibinfo{person}{Daan Wierstra}.}
  \bibinfo{year}{2014}\natexlab{}.
\newblock \showarticletitle{Stochastic Backpropagation and Approximate
  Inference in Deep Generative Models}. In \bibinfo{booktitle}{\emph{Proc. of
  ICML}}.
\newblock


\bibitem[Rogers and Hahn(2010)]%
        {DBLP:journals/jcisd/RogersH10}
\bibfield{author}{\bibinfo{person}{David Rogers} {and} \bibinfo{person}{Mathew
  Hahn}.} \bibinfo{year}{2010}\natexlab{}.
\newblock \showarticletitle{Extended-Connectivity Fingerprints}.
\newblock \bibinfo{journal}{\emph{J. Chem. Inf. Model.}}
  (\bibinfo{year}{2010}).
\newblock


\bibitem[Sacha et~al\mbox{.}(2021)]%
        {Sacha_2021_MEGAN}
\bibfield{author}{\bibinfo{person}{Mikołaj Sacha}, \bibinfo{person}{Mikołaj
  Błaż}, \bibinfo{person}{Piotr Byrski}, \bibinfo{person}{Paweł
  Dąbrowski-Tumański}, \bibinfo{person}{Mikołaj Chromiński},
  \bibinfo{person}{Rafał Loska}, \bibinfo{person}{Paweł
  Włodarczyk-Pruszyński}, {and} \bibinfo{person}{Stanisław Jastrzębski}.}
  \bibinfo{year}{2021}\natexlab{}.
\newblock \showarticletitle{Molecule Edit Graph Attention Network: Modeling
  Chemical Reactions as Sequences of Graph Edits}.
\newblock \bibinfo{journal}{\emph{J. Chem. Inf. Model.}}
  (\bibinfo{year}{2021}).
\newblock


\bibitem[Schneider et~al\mbox{.}(2016)]%
        {schneider2016s}
\bibfield{author}{\bibinfo{person}{Nadine Schneider}, \bibinfo{person}{Nikolaus
  Stiefl}, {and} \bibinfo{person}{Gregory~A Landrum}.}
  \bibinfo{year}{2016}\natexlab{}.
\newblock \showarticletitle{What’s what: The (nearly) definitive guide to
  reaction role assignment}.
\newblock \bibinfo{journal}{\emph{J. Chem. Inf. Model.}}
  (\bibinfo{year}{2016}).
\newblock


\bibitem[Segler and Waller(2017)]%
        {segler2017neural}
\bibfield{author}{\bibinfo{person}{Marwin~HS Segler} {and}
  \bibinfo{person}{Mark~P Waller}.} \bibinfo{year}{2017}\natexlab{}.
\newblock \showarticletitle{Neural-symbolic machine learning for retrosynthesis
  and reaction prediction}.
\newblock \bibinfo{journal}{\emph{Chemistry--A European Journal}}
  (\bibinfo{year}{2017}).
\newblock


\bibitem[Seo et~al\mbox{.}(2021)]%
        {Seo_2021_GTA}
\bibfield{author}{\bibinfo{person}{Seung-Woo Seo}, \bibinfo{person}{You~Young
  Song}, \bibinfo{person}{June~Yong Yang}, \bibinfo{person}{Seohui Bae},
  \bibinfo{person}{Hankook Lee}, \bibinfo{person}{Jinwoo Shin},
  \bibinfo{person}{Sung~Ju Hwang}, {and} \bibinfo{person}{Eunho Yang}.}
  \bibinfo{year}{2021}\natexlab{}.
\newblock \showarticletitle{{GTA}: Graph Truncated Attention for
  Retrosynthesis}. In \bibinfo{booktitle}{\emph{Proc. of AAAI}}.
\newblock


\bibitem[Serban et~al\mbox{.}(2017)]%
        {aaai17}
\bibfield{author}{\bibinfo{person}{Iulian~Vlad Serban},
  \bibinfo{person}{Alessandro Sordoni}, \bibinfo{person}{Ryan Lowe},
  \bibinfo{person}{Laurent Charlin}, \bibinfo{person}{Joelle Pineau},
  \bibinfo{person}{Aaron~C. Courville}, {and} \bibinfo{person}{Yoshua Bengio}.}
  \bibinfo{year}{2017}\natexlab{}.
\newblock \showarticletitle{A Hierarchical Latent Variable Encoder-Decoder
  Model for Generating Dialogues}. In \bibinfo{booktitle}{\emph{Proc. of
  AAAI}}.
\newblock


\bibitem[Shi et~al\mbox{.}(2020)]%
        {g2g}
\bibfield{author}{\bibinfo{person}{Chence Shi}, \bibinfo{person}{Minkai Xu},
  \bibinfo{person}{Hongyu Guo}, \bibinfo{person}{Ming Zhang}, {and}
  \bibinfo{person}{Jian Tang}.} \bibinfo{year}{2020}\natexlab{}.
\newblock \showarticletitle{A Graph to Graphs Framework for Retrosynthesis
  Prediction}. In \bibinfo{booktitle}{\emph{Proc. of ICML}}.
\newblock


\bibitem[Sohn et~al\mbox{.}(2015)]%
        {cvae}
\bibfield{author}{\bibinfo{person}{Kihyuk Sohn}, \bibinfo{person}{Honglak Lee},
  {and} \bibinfo{person}{Xinchen Yan}.} \bibinfo{year}{2015}\natexlab{}.
\newblock \showarticletitle{Learning Structured Output Representation using
  Deep Conditional Generative Models}. In \bibinfo{booktitle}{\emph{Proc. of
  NeurIPS}}.
\newblock


\bibitem[Somnath et~al\mbox{.}(2020)]%
        {Somnath_2020_GRAPHRETRO}
\bibfield{author}{\bibinfo{person}{Vignesh~Ram Somnath},
  \bibinfo{person}{Charlotte Bunne}, \bibinfo{person}{Connor~W. Coley},
  \bibinfo{person}{Andreas Krause}, {and} \bibinfo{person}{Regina Barzilay}.}
  \bibinfo{year}{2020}\natexlab{}.
\newblock \showarticletitle{Learning Graph Models for Template-Free
  Retrosynthesis}.
\newblock \bibinfo{journal}{\emph{ICML Workshop}} (\bibinfo{year}{2020}).
\newblock


\bibitem[S{\o}nderby et~al\mbox{.}(2016)]%
        {DBLP:conf/nips/SonderbyRMSW16}
\bibfield{author}{\bibinfo{person}{Casper~Kaae S{\o}nderby},
  \bibinfo{person}{Tapani Raiko}, \bibinfo{person}{Lars Maal{\o}e},
  \bibinfo{person}{S{\o}ren~Kaae S{\o}nderby}, {and} \bibinfo{person}{Ole
  Winther}.} \bibinfo{year}{2016}\natexlab{}.
\newblock \showarticletitle{Ladder Variational Autoencoders}. In
  \bibinfo{booktitle}{\emph{Proc. of NeurIPS}}.
\newblock


\bibitem[Sun et~al\mbox{.}(2020)]%
        {Sun_2020_EBM}
\bibfield{author}{\bibinfo{person}{Ruoxi Sun}, \bibinfo{person}{Hanjun Dai},
  \bibinfo{person}{Li Li}, \bibinfo{person}{Steven Kearnes}, {and}
  \bibinfo{person}{Bo Dai}.} \bibinfo{year}{2020}\natexlab{}.
\newblock \showarticletitle{Energy-based View of Retrosynthesis}.
\newblock \bibinfo{journal}{\emph{arXiv preprint arXiv:2007.13437}}
  (\bibinfo{year}{2020}).
\newblock


\bibitem[Sutskever et~al\mbox{.}(2014)]%
        {seq_to_seq}
\bibfield{author}{\bibinfo{person}{Ilya Sutskever}, \bibinfo{person}{Oriol
  Vinyals}, {and} \bibinfo{person}{Quoc~V. Le}.}
  \bibinfo{year}{2014}\natexlab{}.
\newblock \showarticletitle{Sequence to Sequence Learning with Neural
  Networks}. In \bibinfo{booktitle}{\emph{Proc. of NIPS}}.
\newblock


\bibitem[Tetko et~al\mbox{.}(2020)]%
        {Tetko_2020_AT}
\bibfield{author}{\bibinfo{person}{I.V. Tetko}, \bibinfo{person}{P. Karpov},
  {and} \bibinfo{person}{R. Van~Deursen}.} \bibinfo{year}{2020}\natexlab{}.
\newblock \showarticletitle{State-of-the-art augmented {NLP} transformer models
  for direct and single-step retro synthesis}.
\newblock \bibinfo{journal}{\emph{Nat. Comm.}} (\bibinfo{year}{2020}).
\newblock


\bibitem[Trost(1991)]%
        {trost1991atom}
\bibfield{author}{\bibinfo{person}{Barry~M Trost}.}
  \bibinfo{year}{1991}\natexlab{}.
\newblock \showarticletitle{The atom economy--a search for synthetic
  efficiency}.
\newblock \bibinfo{journal}{\emph{Science}} (\bibinfo{year}{1991}).
\newblock


\bibitem[Tu and Coley(2021)]%
        {g2s}
\bibfield{author}{\bibinfo{person}{Zhengkai Tu} {and} \bibinfo{person}{Connor~W
  Coley}.} \bibinfo{year}{2021}\natexlab{}.
\newblock \showarticletitle{Permutation invariant graph-to-sequence model for
  template-free retrosynthesis and reaction prediction}.
\newblock \bibinfo{journal}{\emph{arXiv preprint arXiv:2110.09681}}
  (\bibinfo{year}{2021}).
\newblock


\bibitem[Vaswani et~al\mbox{.}(2017)]%
        {vaswani2017attention}
\bibfield{author}{\bibinfo{person}{Ashish Vaswani}, \bibinfo{person}{Noam
  Shazeer}, \bibinfo{person}{Niki Parmar}, \bibinfo{person}{Jakob Uszkoreit},
  \bibinfo{person}{Llion Jones}, \bibinfo{person}{Aidan~N Gomez},
  \bibinfo{person}{{\L}ukasz Kaiser}, {and} \bibinfo{person}{Illia
  Polosukhin}.} \bibinfo{year}{2017}\natexlab{}.
\newblock \showarticletitle{Attention is all you need}. In
  \bibinfo{booktitle}{\emph{Proc. of NeurIPS}}.
\newblock


\bibitem[Wang et~al\mbox{.}(2021)]%
        {Wang_2021_RetroPRIME}
\bibfield{author}{\bibinfo{person}{Xiaorui Wang}, \bibinfo{person}{Yuquan Li},
  \bibinfo{person}{Jiezhong Qiu}, \bibinfo{person}{Guangyong Chen},
  \bibinfo{person}{Huanxiang Liu}, \bibinfo{person}{Benben Liao},
  \bibinfo{person}{Chang-Yu Hsieh}, {and} \bibinfo{person}{Xiaojun Yao}.}
  \bibinfo{year}{2021}\natexlab{}.
\newblock \showarticletitle{RetroPrime: A Diverse, plausible and
  Transformer-based method for Single-Step retrosynthesis predictions}.
\newblock \bibinfo{journal}{\emph{Chemical Engineering Journal}}
  (\bibinfo{year}{2021}).
\newblock


\bibitem[Weininger(1988)]%
        {weininger1988smiles}
\bibfield{author}{\bibinfo{person}{David Weininger}.}
  \bibinfo{year}{1988}\natexlab{}.
\newblock \showarticletitle{SMILES, a chemical language and information system.
  1. Introduction to methodology and encoding rules}.
\newblock \bibinfo{journal}{\emph{J. Chem. Inf. Model.}}
  (\bibinfo{year}{1988}).
\newblock


\bibitem[Yan et~al\mbox{.}(2020)]%
        {retrox}
\bibfield{author}{\bibinfo{person}{Chaochao Yan}, \bibinfo{person}{Qianggang
  Ding}, \bibinfo{person}{Peilin Zhao}, \bibinfo{person}{Shuangjia Zheng},
  \bibinfo{person}{Jinyu Yang}, \bibinfo{person}{Yang Yu}, {and}
  \bibinfo{person}{Junzhou Huang}.} \bibinfo{year}{2020}\natexlab{}.
\newblock \showarticletitle{RetroXpert: Decompose Retrosynthesis Prediction
  Like {A} Chemist}. In \bibinfo{booktitle}{\emph{Proc. of NeurIPS}}.
\newblock


\bibitem[Yang et~al\mbox{.}(2019b)]%
        {dmpnn}
\bibfield{author}{\bibinfo{person}{Kevin Yang}, \bibinfo{person}{Kyle Swanson},
  \bibinfo{person}{Wengong Jin}, \bibinfo{person}{Connor~W. Coley},
  \bibinfo{person}{Philipp Eiden}, \bibinfo{person}{Hua Gao},
  \bibinfo{person}{Angel Guzman{-}Perez}, \bibinfo{person}{Timothy Hopper},
  \bibinfo{person}{Brian Kelley}, \bibinfo{person}{Miriam Mathea},
  \bibinfo{person}{Andrew Palmer}, \bibinfo{person}{Volker Settels},
  \bibinfo{person}{Tommi~S. Jaakkola}, \bibinfo{person}{Klavs~F. Jensen}, {and}
  \bibinfo{person}{Regina Barzilay}.} \bibinfo{year}{2019}\natexlab{b}.
\newblock \showarticletitle{Analyzing Learned Molecular Representations for
  Property Prediction}.
\newblock \bibinfo{journal}{\emph{J. Chem. Inf. Model.}}
  (\bibinfo{year}{2019}).
\newblock


\bibitem[Yang et~al\mbox{.}(2019a)]%
        {yang2019molecular}
\bibfield{author}{\bibinfo{person}{Qingyi Yang}, \bibinfo{person}{Vishnu
  Sresht}, \bibinfo{person}{Peter Bolgar}, \bibinfo{person}{Xinjun Hou},
  \bibinfo{person}{Jacquelyn~L Klug-McLeod}, \bibinfo{person}{Christopher~R
  Butler}, {et~al\mbox{.}}} \bibinfo{year}{2019}\natexlab{a}.
\newblock \showarticletitle{Molecular Transformer unifies reaction prediction
  and retrosynthesis across pharma chemical space}.
\newblock \bibinfo{journal}{\emph{Chemical Communications}}
  (\bibinfo{year}{2019}).
\newblock


\bibitem[Yu et~al\mbox{.}(2020)]%
        {aaai20}
\bibfield{author}{\bibinfo{person}{Meng{-}Hsuan Yu}, \bibinfo{person}{Juntao
  Li}, \bibinfo{person}{Danyang Liu}, \bibinfo{person}{Bo Tang},
  \bibinfo{person}{Haisong Zhang}, \bibinfo{person}{Dongyan Zhao}, {and}
  \bibinfo{person}{Rui Yan}.} \bibinfo{year}{2020}\natexlab{}.
\newblock \showarticletitle{Draft and Edit: Automatic Storytelling Through
  Multi-Pass Hierarchical Conditional Variational Autoencoder}. In
  \bibinfo{booktitle}{\emph{Proc. of AAAI}}.
\newblock


\bibitem[Yuan et~al\mbox{.}(2018)]%
        {yuan2018retrosynthesis}
\bibfield{author}{\bibinfo{person}{Shuai Yuan}, \bibinfo{person}{Jun-Sheng
  Qin}, \bibinfo{person}{Jialuo Li}, \bibinfo{person}{Lan Huang},
  \bibinfo{person}{Liang Feng}, \bibinfo{person}{Yu Fang},
  \bibinfo{person}{Christina Lollar}, \bibinfo{person}{Jiandong Pang},
  \bibinfo{person}{Liangliang Zhang}, \bibinfo{person}{Di Sun},
  {et~al\mbox{.}}} \bibinfo{year}{2018}\natexlab{}.
\newblock \showarticletitle{Retrosynthesis of multi-component metal-organic
  frameworks}.
\newblock \bibinfo{journal}{\emph{Nat. Comm.}} (\bibinfo{year}{2018}).
\newblock


\bibitem[Zhao et~al\mbox{.}(2017)]%
        {acl17}
\bibfield{author}{\bibinfo{person}{Tiancheng Zhao}, \bibinfo{person}{Ran Zhao},
  {and} \bibinfo{person}{Maxine Esk{\'{e}}nazi}.}
  \bibinfo{year}{2017}\natexlab{}.
\newblock \showarticletitle{Learning Discourse-level Diversity for Neural
  Dialog Models using Conditional Variational Autoencoders}. In
  \bibinfo{booktitle}{\emph{Proc. of ACL}}.
\newblock


\bibitem[Zheng et~al\mbox{.}(2020)]%
        {Zheng_2020_SCROP}
\bibfield{author}{\bibinfo{person}{Shuangjia Zheng}, \bibinfo{person}{Jiahua
  Rao}, \bibinfo{person}{Zhongyue Zhang}, \bibinfo{person}{Jun Xu}, {and}
  \bibinfo{person}{Yuedong Yang}.} \bibinfo{year}{2020}\natexlab{}.
\newblock \showarticletitle{Predicting Retrosynthetic Reactions Using
  Self-Corrected Transformer Neural Networks}.
\newblock \bibinfo{journal}{\emph{J. Chem. Inf. Model.}}
  (\bibinfo{year}{2020}).
\newblock


\bibitem[Zhou and Wang(2018)]%
        {mojitalk}
\bibfield{author}{\bibinfo{person}{Xianda Zhou} {and}
  \bibinfo{person}{William~Yang Wang}.} \bibinfo{year}{2018}\natexlab{}.
\newblock \showarticletitle{MojiTalk: Generating Emotional Responses at Scale}.
  In \bibinfo{booktitle}{\emph{Proc. of ACL}}.
\newblock


\bibitem[Zhu et~al\mbox{.}(2021)]%
        {zhu2021dual}
\bibfield{author}{\bibinfo{person}{Jinhua Zhu}, \bibinfo{person}{Yingce Xia},
  \bibinfo{person}{Tao Qin}, \bibinfo{person}{Wengang Zhou},
  \bibinfo{person}{Houqiang Li}, {and} \bibinfo{person}{Tie-Yan Liu}.}
  \bibinfo{year}{2021}\natexlab{}.
\newblock \showarticletitle{Dual-view Molecule Pre-training}.
\newblock \bibinfo{journal}{\emph{arXiv preprint arXiv:2106.10234}}
  (\bibinfo{year}{2021}).
\newblock


\end{thebibliography}

% \newpage
% \appendix
% \input{section/appendix}

\end{document}